\documentclass[12pt,letterpaper]{article}
\usepackage{jheppub}
\usepackage{color}

\usepackage{epsfig,multirow,subfigure} 
\usepackage{amscd}
\usepackage[matrix,arrow,curve]{xy}
\usepackage{verbatim}
\usepackage{latexsym}
\usepackage{amsfonts,amsthm,amsmath,mathtools}

\newcommand{\be}{\begin{equation}}
\newcommand{\ee}{\end{equation}}
\newcommand{\bea}{\begin{eqnarray}}
\newcommand{\eea}{\end{eqnarray}}
\newcommand{\bef}{\begin{figure}}
\newcommand{\eef}{\end{figure}}

\newcommand{\non}{\nonumber}

\title{\centering
Spin$(7)$ duality for $\mathcal{N}=1$ CS-matter theories
}

\author[a]{A. Amariti,}
\author[b,c]{D. Forcella}

\affiliation[a]{
  Laboratoire de Physique Th\'eorique de l'\'Ecole Normale Sup\'erieure and\\
  Istitute de Physique Th\'eorique Philippe Meyer, 24 Rue Lhomond,
  Paris 75005, France}
\affiliation[b]{
Physique Th\'eorique et Math\'ematique and International Solvay Institutes\\
Universit\'e Libre de Bruxelles, C.P. 231, 1050 Bruxelles, Belgium}
\affiliation[c]{ Laboratoire de Physique Th\'eorique et Hautes Energies, Universit\'e Pierre et Marie Curie,\\
4 Place Jussieu, 75252 Paris Cedex 05, France}
\emailAdd{amariti@lpt.ens.fr}
\emailAdd{dforcell@ulb.ac.be}

\abstract{In this paper we propose a duality for non-holomorphic
$\mathcal{N}=1$ CS-matter theories living on M2 branes probing
Spin$(7)$ cones.  We call this duality Spin$(7)$ duality.  Two
theories are named Spin$(7)$ dual if they have the same 
moduli space: a real Spin$(7)$ cone with base a weak G$_2$ manifold,
and they are hence holographic dual to the same AdS$_4$ $\times$ G$_2$
M theory solution.  We provide a systematic way to generate these
dualities, derived by combining toric duality for $\mathcal{N}=2$
CS-matter theories and generalized non-holomorphic orientifold
projections to $\mathcal{N}=1$.  Brane construction, AdS/CFT
correspondence, and the computation of the moduli space support our
proposal at the classical level and provide some arguments at the
quantum strong coupling regime. The relation with Seiberg-like duality
is also analyzed.}

\begin{document}

\maketitle
%
%
%
%
%
%
%
%
%
%
%
\section{Introduction}
\label{sec:Intro}

Strongly coupled systems are interesting both from phenomenological
and theoretical perspectives.  However, understanding their dynamics
is usually quite difficult. An interesting strategy to explore such
phenomena consists of looking for an alternative, weakly coupled,
description of the same system, the so called ``dual description''.
Dualities have been discovered and studied in many contexts and they
provided a deep insight in strongly coupled physics.  Supersymmetric
field theories are an useful laboratory to explore duality maps.
Seiberg duality for SQCD \cite{Seiberg:1994pq} in four dimensions and
its generalizations to $\mathcal{N}=2$ three dimensional field
theories \cite{
  Aharony:1997gp,Karch:1997ux,Aharony:2008gk,Giveon:2008zn,Amariti:2009rb,
  Benini:2011mf, Closset:2012eq} are examples of this map.  Another
well known duality for three dimensional field theories is the
AdS$_4/$CFT$_3$ correspondence
\cite{Aharony:2008ug,Hanany:2008cd,Martelli:2008si} that relates
Chern-Simons (CS) matter theories to M theory AdS$_4$ solutions.  In
this case it has been shown that there are different UV field theory
descriptions of the IR theory living on M2 branes probing the same
toric Calabi-Yau four dimensional cone CY$_4$
\cite{Amariti:2009rb,Franco:2009sp,Davey:2009sr,Benini:2009qs,Dwivedi:2014awa}.
This phenomenon has been named toric duality and it is the three
dimensional extension of the previously discovered toric duality for
the four dimensional field theories living on D3-branes probing
Calabi-Yau three dimensional cones CY$_3$
\cite{Feng:2000mi,Feng:2001xr,Feng:2002zw,Forcella:2008ng}. For four
dimensional field theories toric duality coincides with Seiberg
duality \cite{Beasley:2001zp,Feng:2001bn}.

It was shown in \cite{Amariti:2009rb} that, for some classes of
theories, toric duality for M2 branes is a generalization of the
$\mathcal{N}=2$ Seiberg-like duality of \cite{Giveon:2008zn}.
$\mathcal{N}=2$ CS-matter theories can be further reduced to
$\mathcal{N}=1$ CS-matter theories \footnote{See
  \cite{Mauri:2008ai,Ooguri:2008dk,Gaiotto:2009mv,Bobev:2009ms} for
  some recent analysis of $\mathcal{N}=1$ theories in three
  dimensions} living on stacks of M2 branes at certain conical
singularities \cite{Forcella:2009jj}: the so called Spin$(7)$ cones.
These theories can be obtained with a generalized orientifold
projection from parents $\mathcal{N}=2$ holomorphic theories
describing stacks of $N$ M2 branes probing the tip of toric CY$_4$
cones.  In the geometric language this projection corresponds to the
quotient done by an anti-holomorphic involution on the CY$_4$, that
breaks the $SU(4)$ holonomy to Spin$(7)$ \footnote{ We refer the
  reader to \cite{Bonetti:2013fma,Bonetti:2013nka} where the reduction
  of M-theory on Spin$(7)$ manifolds constructed by this method has
  been considered too.}  \cite{Joyce:1999nk,Forcella:2009jj}.

Inspired by the $\mathcal{N}=2$ case one may ask if there are
extensions of toric (and of Seiberg-like) duality to the
$\mathcal{N}=1$ case \footnote{We refer the reader to
  \cite{Armoni:2009vv} for another proposal of Seiberg-like duality in
  $\mathcal{N}=1$ theories.}.  In this paper we use a geometrical
approach and define a Spin$(7)$ duality in analogy with the toric
duality of the toric CY$_4$ case.  Namely we say that two
$\mathcal{N}=1$ CS-matter theories are Spin$(7)$ dual if they have the
same classical moduli space for one regular M2 brane and if it
coincides with the Spin$(7)$ cone of the dual geometry.  We provide a
general picture to generate $\mathcal{N}=1$ Spin$(7)$ dual pairs
obtained from parent toric dual $\mathcal{N}=2$ theories.  Some
control on these dualities beyond the classical level is provided by
the existence of the same AdS$_4$ dual geometry for both the dual CFTs
and by planar equivalence.

In some cases the orientifold projects the $\mathcal{N}=2$ theory to
$\mathcal{N}=1$ theories with only unitary groups.  In these cases we
argue that the Spin$(7)$ duality is also an $\mathcal{N}=1$ three
dimensional Seiberg-like duality.  Indeed it corresponds to move
$\mathcal{N}=1$ branes in the Hanany-Witten \cite{Hanany:1996ie}
projected setup.

The paper is organized as follows. In section \ref{sec:aoi} we review
the main aspects of the projection of the CY$_4$ to Spin$(7)$ and its
interpretation in terms of an orientifold.  In section
\ref{sec:general} we state the main claim of the paper about the
$\mathcal{N}=1$ Spin$(7)$ duality and explain the general idea behind
it.  In section \ref{sec:examples} we provide some examples of dual
pairs and give some checks about the validity of the duality.  In
section \ref{sec:geno} we show examples where the Spin$(7)$ duality
can be regarded as a Seiberg-like duality.  In section
\ref{sec:holodual} we discuss the extension of $\mathcal{N}=1$ Seiberg
like duality to more general models.  In section \ref{sec:conclusions}
we conclude.  To complete the paper we provide also two appendices. In
appendix \ref{APPA} we explain the projection of the $\mathcal{N}=2$
superspace to $\mathcal{N}=1$ while in appendix \ref{APPB} we present
the $\mathcal{N}=1$ superconformal algebra.
%
%
%
%
%
%
%
%
%
%
%
%
%
%
%
%
%
%
\section{From $\mathcal{N}=2$ CY$_4$ to $\mathcal{N}=1$
Spin$(7)$}
\label{sec:aoi}
In this section we briefly review the non-holomorphic orbifold of the
CY$_4$ geometry that we will use in the rest of the paper \cite{Joyce:1999nk}, 
and we will provide a short discussion of the associated
 orientifold projection in field theory\footnote{In
  the next section we will report some more details on the field
  theory.}.  An interesting class of $\mathcal{N}=2$ SCFTs 
\cite{Hanany:2008cd,Martelli:2008si} describes the low energy dynamics of a
stack of $N$ M2 branes at the tip of a non compact eight-dimensional
CY$_4$ real cone: C$(H_7)$, where $H_7$ is a seven dimensional compact
Sasaki-Einstein manifold at the base of the cone.

The field theory is a quiver gauge theory.  A quiver is a graph with
nodes connected by arrows.  Each node represents a gauge factor
$U(N_i)$. There are also matter fields, represented by oriented
arrows.  Arrows with the tip and the tail on the same node are fields in the
adjoint representation of the gauge group, arrows connecting the
$i$-th with the $j$-th node are associated to fields in the bifundamental
representation.  In the Lagrangian each $U(N_i)$ factor has  CS
action with integer level $k_i$, and no Yang-Mills (YM) action. 
From now on we will keep track of the CS level and the rank of
the gauge group factor by using the notation: $U(N_i)_{k_i}$.

These field theories are dual, in the gauge/gravity correspondence, to
M-theory on the AdS$_4 \times H_7$ background.  In this paper we
consider a particular projection of this theory that breaks the four
real supercharges down to two real supercharges\footnote{In M-theory
  the background $\mathbb{R}^{1,2} \times C$ preserves four real
  supercharges ($\mathcal{N}=2$ susy in three dimension), if $C$ is a
  CY$_4$ manifold, or two real supercharges ($\mathcal{N}=1$ susy in
  three dimension), if $C$ is a Spin$(7)$ manifold
  \cite{Morrison:1998cs,Acharya:1998db}.}.  The resulting theory is still a
superconformal CS-matter theory, like before, but with only
$\mathcal{N}=1$ supersymmetry in three dimension.  Moreover it does
not have holomorphic properties: fields and superpotential are
real.  It describes the low energy dynamics of $N$ M2 branes living at
the tip of a Spin$(7)$ cone: $C(G_2)$, where G$_2$ is a seven
dimensional compact weak G$_2$ manifold.  These theories are dual, in
the gauge/gravity correspondence to M-theory on the AdS$_4 \times G_2$
background.

On the geometry the projection is obtained by modding the original
CY$_4$ by the action of an anti-holomorphic involution $\Theta$
\cite{Joyce:1999nk}.  This geometric procedure is implemented in field
theory by projecting the lagrangian  using an 
  orientifold projection \cite{Forcella:2009jj} as we will review in
the rest of this section and in the following section.

A CY$_4$ has a Kahler $(1,1)$ form $J$ and a holomorphic $(4,0)$ form
$\omega$, that are left invariant by the holonomy group of the
manifold: $SU(4)$.  Following \cite{Joyce:1999nk} we use the action
of an anti-holomorphic involution $\Theta$ to define a Spin$(7)$
manifold.  $\Theta$ acts on $J$ and $\omega$ as $\Theta:\omega
\rightarrow \overline \omega$, and $\Theta:J\rightarrow -J$, and it
breaks the $SU(4)$ holonomy to Spin$(7)$.  Using the defining forms of
the CY$_4$ it is indeed possible to construct a closed self dual four
form
\begin{equation}
\Omega_4 =\frac{1}{2}J \wedge J+Re(\omega)
\end{equation}
that is left invariant under the action of $\Theta$ and hence defines
a Spin$(7)$ manifold \cite{Joyce:1999nk}.  In the field theory it is
possible to interpret a class of these quotients as an orientifold
\cite{Sen:1996zq,Gopakumar:1996mu,Majumder:2001dx,Partouche:2000uq,Forcella:2009jj}.
Because there are no open strings in M-theory it is easier to define
its action by looking at the type IIA limit.  Indeed the CY$_4$ cone
$Y$ that we consider can be written as a double fibration of a CY$_3$
$Z$, over a real line, parameterized by the real coordinate $\sigma$,
and a circle, parameterized by an angle $\psi$
\cite{Hanany:2008cd,Martelli:2008si,Aganagic:2009zk,HananyUnp,Closset:2012ep}.  The
angle $\psi$ parameterizes the M-theory circle while $\sigma$ is the
expectation value of a particular combination of the D terms in field
theory. In the type IIA limit one describes the worldvolume theory of
D2 branes probing a seven dimensional manifold given by $Z$ fibered
over a line.  The four form $\omega$ is locally
\begin{equation}
\omega \sim
f(z_i)
d z_1 \wedge d z_2 \wedge d z_3 \wedge ( d\sigma + i d \psi)
\end{equation}
where $z_i$ are the holomorphic coordinates of $Z$ and $f(z_i)$ is a
holomorphic function.  We choose an anti-holomorphic involution
$\Theta$ that acts on the M-theory circle as $\Theta:\psi\rightarrow
-\psi$, and that leaves invariant the coordinate $\sigma$. This class
of quotients in M-theory can then be interpreted as an orientifold
projection
\cite{Sen:1996zq,Gopakumar:1996mu,Majumder:2001dx,Partouche:2000uq,Forcella:2009jj}.
One then concludes that the field theory living on the M2 branes at the
tip of $Y/\Theta$ geometry is the IR strong coupling limit in M-theory
of the $\mathcal{N}=1$ orientifold theory living on a stack of $N$ D2
branes in type IIA \cite{Forcella:2009jj}.  From now on we refer to
the $\mathcal{N}=2$ theories as the ``parent theories'', while we
refer to the $\mathcal{N}=1$ theories as the ``projected theories''.
%
%
%
%
%
%
%
%
%
%
%
%
%
%
%
%
%
%
\section{Spin$(7)$ duality: our strategy}
\label{sec:general}

In this section we discuss our approach to generate and check
Spin$(7)$ dualities between $\mathcal{N}=1$ three dimensional
CS-matter theories with gravity duals.

First we give some general remarks of the Spin$(7)$ duality that we
are proposing.  Two UV $\mathcal{N}=1$ field theories are Spin$(7)$
dual if their moduli spaces coincide and they are equivalent to the
Spin$(7)$ cone probed by one M2 brane.  This duality is the analogous
of the toric duality for $\mathcal{N}=2$ theories living at the tip of
toric CY$_4$ cones.  There are some important differences between the
two dualities.  First in the $\mathcal{N}=2$ case the gauge theory
that lives on an M2 brane is abelian, while in the $\mathcal{N}=1$
case the theory for a single M2 is usually non-abelian.  Second, both
toric and Spin$(7)$ duality are classical dualities.  In the
$\mathcal{N}=2$ case the duality is valid also at quantum level. The
$\mathcal{N}=1$ theories are not holomorphic and one may expect
quantum corrections.  Anyway the underlining AdS/CFT duality provides
some arguments supporting the duality also in the quantum strongly
coupled regime. Further studies are however required to understand the
quantum properties of the proposed Spin$(7)$ duality, and we leave
them for future works. A last important remark concerns the relation
between Spin$(7)$ duality and Seiberg-like duality.  For
$\mathcal{N}=2$ three dimensional CS-matter theories it has been shown
that, for a particular class of theories, the so called
$\widetilde{L^{aba}_{k}}$ models, some toric dualities are actually
Seiberg-like dualities \cite{Amariti:2009rb}.  In this paper we will
discuss some cases in which also the Spin$(7)$ duality is a
Seiberg-like duality.

In the following we provide a step by step illustration of our
strategy to obtain $\mathcal{N}=1$ pairs, and to check the validity of
the Spin$(7)$ duality.  We start by introducing in some details the
$\mathcal{N}=2$ parent theories living on $N$ M2 branes at the tip of
a CY$_4$ cone, and discuss their moduli space.  A discussion on the
orientifold projection to $\mathcal{N}=1$ in field theory
follows. Then we explain our general strategy to obtain the moduli
space of $\mathcal{N}=1$ field theories and to match the moduli space
and the geometry of $\mathcal{N}=1$ field theory dual pairs.  We
conclude with a discussion on the relation between Spin$(7)$ duality
and Seiberg-like duality.  More details could be found in
\cite{Hanany:2008cd,Martelli:2008si,Amariti:2009rb,Forcella:2009jj}.
From now on we will refer to the $\mathcal{N}=2$ field theory as the "parent 
theory", while we will call the $\mathcal{N}=1$ theory the "projected theory".

\subsection*{The $\widetilde{L^{aba}}_{k_i}$ $\mathcal{N}=2$ CS-matter
  theories}

The $\mathcal{N}=2$ parents theories we consider are three dimensional
extensions of $L^{aba}$ four dimensional quiver gauge theories
\cite{Benvenuti:2005ja,Butti:2005sw,Franco:2005sm}, introduced in
\cite{Martelli:2008si,Hanany:2008cd}. They are CS-matter theories with
a product of $U(N_i)$ gauge groups and CS levels $k_i$, with
$i=1,...a+b$, with pairs of bifundamental-antibifundamental connecting
each pair of consecutive $U(N_i)$ and possibly adjoint fields.  Every
field appears twice in the superpotential with opposite sign, such
that every $F$-term is an equality between two monomials with the same
sign. From now on we will refer to these theories as
$\widetilde{L^{aba}}_{k_i}$. Examples of the  $\widetilde{L^{aba}}_{k_i}$ quivers
are given  in figure \ref{fig:L2221} and \ref{fig:L444}. 

These are the low energy theories living on M2 branes at particular
CY$_4$ singularities that are the double fibration of the $L^{aba}$
CY$_3$ singularity over a segment parameterized by $\sigma$ and a
circle parameterized by $\psi$. In the UV they have a simple type IIB
description in terms of branes
\cite{Hanany:1996ie,Kitao:1998mf,Bergman:1999na,Aharony:2008ug}.

\subsubsection*{Brane setup and dualities}
The $\widetilde{L^{aba}}_{k_i}$ theories can be engineered as a stack
of D3 branes on a circle ending on a set of $(1,p_i)$ five-branes,
where $i=1,\dots,a+b$: $N_i$ D3s for every interval between a
$(1,p_i)$ and a $(1,p_{i+1})$ five-brane. This construction corresponds
to a circular quiver with $a+b$ gauge groups and a pair
bifundamental-antibifundamental connecting each pair of consecutive
nodes that are actually the type IIB strings stretching through the
$i$-th five-branes.  
The $N$ D3 branes are extended along the directions $(x_0,x_1,x_2)$
and the direction $x_6$ compactified on a circle.
The NS5 and the D5 branes, that recombine into the five-branes, are
divided in two sets.  In the first case one NS is extended along
$(x_0,x_1,x_2,x_3,x_4,x_5)$ and the corresponding $p_i$ D5 are
extended along $(x_0,x_1,x_2,x_4,x_5,x_7)$.  In the second case one NS
is extend-end along $(x_0,x_1,x_2,x_3,x_8,x_9)$ and the corresponding
$p_i$ D5 are extended along $(x_0,x_1,x_2,x_7,x_8,x_9)$. There are $a$
$(1,p_i)$ five-branes of the first type and $b$ five-branes of the
second type.  The SCFT lives in the $(x_0,x_1,x_2)$ directions common
to all the branes.  
The NS branes and the corresponding D5 branes get deformed in $(1,p_i)$ 
five-branes at angles $\tan \theta_i \simeq p_i$.
The Chern-Simons levels are associated with the relative angle of the branes in
the $(3,7)$ directions, they are $k_i=p_i-p_{i+1}$, such that $\sum_i k_i =0$. 
When the $(1,p_i)$ and the $(1,p_{i+1})$
five-branes are parallel there is a massless adjoint field associated to the
$i$-th gauge group.  In the minimal phase there are $b-a$ nodes with
an adjoint fields and $2a$ nodes without the adjoint.

By exchanging two consecutive (non parallel) five-branes one has a
local transformation on the quiver, that corresponds to a Seiberg-like
duality in field theory. If this action is performed on the $i$-th
gauge group we have the transformation \cite{Amariti:2009rb}
\begin{eqnarray}
\label{N2r}
U(N)_{k_{i-1}} &\rightarrow& U(N)_{k_i+k_{i-1}} \nonumber \\ 
U(N)_{k_i} &\rightarrow& U(N+|k_i|)_{-k_i}  \\ 
U(N)_{k_{i+1}} &\rightarrow& U(N)_{k_i+k_{i+1}}  \nonumber 
\end{eqnarray}
It is possible to demonstrate in full generality that this local
transformation preserves the moduli space \cite{Amariti:2009rb}:
CY$_4$ moduli space associated to the same dual supergravity background.

\subsubsection*{Moduli Space}

  The moduli space of these theories is the set of
values of the scalar fields that solve the zero condition for the
bosonic potential.  This boils down to solve the following set of
equations.
\begin{eqnarray}
\label{VB0}
\partial_{X_{ab}} W &=& 0 \nonumber \\
D_a(X) &=& \frac{k_a \sigma_a}{2 \pi}\nonumber \\
\sigma_a X_{ab} - X_{ab} \sigma_b &=& 0\nonumber \\
\end{eqnarray}
where $W$ is the superpotential, $X_{ab}$ are scalar components of the
bifundamental fields between the $U(N_a)$ and the $U(N_b)$ factor of
the gauge group\footnote{With some abuse of notation we will often use
  the same symbol: $X_{ab}$ to refer both to the superfield or to its
  lowest scalar component. We hope that the reader will not get
  confused. What we meant should be clear from the context.}, $D_a(X)$
is a real function of the bifundamental fields that corresponds to the
usual D-terms, and $\sigma_a$ are the real scalar components of the
vector multiplet for the $U(N_a)$ factors.

For $N$ M2 branes at the tip of the cone, without fractional branes,
we have: $N_a=N_b=N$.  The moduli space is then simply the $N$-times
symmetric product of the moduli space for one brane.  For one regular
M2 brane, the gauge group is simply $U(1)^G$. The moduli space is
found by imposing the set of three equations in (\ref{VB0}) and by
quotienting by the appropriate gauge group factors. It is important to
notice that in the abelian case the third equation in (\ref{VB0})
simply imposes: $\sigma_a=\sigma$, while one of the D-term equations
is redundant, because $\sum_a k_a = \sum_a D_a(X)=0$.  Then we are
left with $G-1$ linearly independent equations.  One of these
equations  can be written along the direction of
the CS levels and it fixes the value of $\sigma_a= \sigma$, while the
remaining $G-2$ are orthogonal to this direction and equate the $G-2$
linear combinations of D terms to zero.  We should then quotient by
the associated $G-2$ $U(1)$ factors, while the $U(1)$ corresponding to
the D-term orthogonal to the CS is broken to
$\mathbb{Z}_{gcd{\{k_a\}}}=\mathbb{Z}_k$ and only imposes an
additional discrete quotient.  The moduli space of an $\mathcal{N}=2$
CS-matter theory is then in general a $\mathbb{Z}_k$ quotient of a
CY$_4$ cone $Y$, where $k$ is the maximum common divisor of the CS
levels \cite{Martelli:2008si,Hanany:2008cd}.

The analysis of the moduli space of the dual pairs generated using the
transformation (\ref{N2r}) for the $\widetilde{L^{aba}}_{k_i}$
theories was done in \cite{Amariti:2009rb} and it was shown that these
models have the same moduli space  and are toric dual.  The main claim of this paper is
that similar dualities exist in the $\mathcal{N}=1$ case, when the
dual geometry is described by a Spin$(7)$ manifold obtained as
explained in section \ref{sec:aoi}. To support this claim we 
provide a coherent geometrical and brane-orientifold construction. The
CY$_4$ moduli space of two toric dual $\mathcal{N}=2$ theories is projected
on the same Spin$(7)$.

We first show how to compute in the $\mathcal{N}=1$ case the moduli
space for a single M2 brane probing a Spin$(7)$ cone.  This is the
non-holomorphic quotient of the original CY$_4$.  We then check that
the two $\mathcal{N}=1$ theories, claimed to be Spin $(7)$ dual,
obtained by projecting the parent $\mathcal{N}=2$ theories, have the
same moduli space.

\subsection*{Field theory projection to $\mathcal{N}=1$}

As explained in section \ref{sec:aoi} the Spin$(7)$ cone is obtained
by quotienting the CY$_4$ $Y$ by the anti-holomorphic involution
$\Theta$. This corresponds to a real orbifold of $Y$ in M-theory and
it acts as an orientifold on the dual field theory.
In this subsection we briefly discuss the action of the projection on
the field theory lagrangian, while in the next subsection we show that
the moduli space of the projected theory is actually the Spin$(7)$
geometry \cite{Forcella:2009jj}.

There are two interesting classes of orientifold projections.  In the
first class the orientifold action identifies the gauge groups with
themselves, projecting the unitary $U(N)$ groups of the
$\mathcal{N}=2$ parent theory to orthogonal $O(2N)$ and/or symplectic
$SP(2N)$ groups in the $\mathcal{N}=1$ projected theory\footnote{Please observe that the standard orientifold procedure implies that
we should double the ranks of the gauge groups and the CS-levels before quotienting the theory. Here
  we use the convention $SP(2)_k=SU(2)_{2k}$ for the symplectic cases.}. In
the second class the orientifold action instead identifies pairs of
$U(N)$ gauge group factors of the $\mathcal{N}=2$ parent theory
projecting them to a single $U(2N)$ group in the $\mathcal{N}=1$
projected theory.  It is important to underline that the orientifold
acts in general as an anti-holomorphic involution on the matter fields
in the lagrangian and it breaks the holomorphic structure of the
$\mathcal{N}=2$ theory, preserving only $\mathcal{N}=1$ supersymmetry.

In the first class, where the projection identifies the $a$-th group
with itself, the orientifold acts on the gauge and matter fields as:
\begin{eqnarray}
\label{tr1}
A_\mu^a &\rightarrow& - \Omega_a (A_\mu^a)^T \Omega_a^{-1} \nonumber \\
X_{ab} &\rightarrow& \Omega_a X_{ab}^* \Omega_a^{-1} \nonumber \\
\sigma_a &\rightarrow&  \Omega_a \sigma_a^T \Omega_a^{-1} \nonumber \\
D_a &\rightarrow&  \Omega_a D_a^T \Omega_a^{-1} 
\end{eqnarray} 
where $\Omega_a$ could be either the identity or the symplectic
matrix. When $\Omega_a=I_{2N}$ it projects the unitary to an
orthogonal group, if instead $\Omega_a= J_{2N}$ it projects the unitary to a
symplectic group.

In the second class, where instead the projection identifies pairs of
groups, $a_i \leftrightarrow b_i$, the orientifold acts on the gauge
and matter fields as:
\begin{eqnarray}
\label{tr2}
A_\mu^a &\rightarrow& - \Omega_{ab} (A_\mu^b)^T \Omega_{ab}^{-1} \nonumber \\
X_{a_1 a_2} &\rightarrow& \Omega_{a_1 b_1} X_{b_1 b_2}^* \Omega_{a_2 b_2}^{-1} 
\nonumber \\
\sigma_a &\rightarrow&  \Omega_{ab} \sigma_a^T \Omega_{ab}^{-1} \nonumber \\
D_a &\rightarrow&  \Omega_{ab} D_b^T \Omega_{ab}^{-1}
\end{eqnarray}
where, as before, the $\Omega_{ab}$ matrix could be either the
identity or the symplectic matrix.

In both cases, because $A$ and $\sigma$ have different transformation
rules, the $\mathcal{N}=2$ vector multiplet is broken to the sum of
the $\mathcal{N}=1$ vector multiplet and the real
$\mathcal{N}=1$ matter multiplet.  Moreover it is manifest in
(\ref{tr1}) and (\ref{tr2}) that the involution breaks the holomorphic
structure of the superpotential.  The details on the $\mathcal{N}=1$
lagrangian are reported in appendix \ref{APPA}.

It is maybe important to remind that the action for which we quotient
the $\mathcal{N}=2$ theory is a symmetry of the theory itself.

\subsubsection*{The moduli space of $\mathcal{N}=1$  theories and Spin$(7)$ duality}

As discussed above we have a Spin$(7)$ duality if the proposed pair of
field  theories have  the  same moduli  space  for one  M2 brane:  the
Spin$(7)$ cone obtained as  the non-holomorphic quotient of the CY$_4$
cone moduli space of the parent theories.  Here we sketch our strategy
to compute the moduli space and verify the Spin$(7)$ duality.

The moduli space for one M2 brane is obtained by setting $N=1$ in all
the gauge group factors.
It is important to underline that finding the $\mathcal{N}=1$ moduli
space for one M2 brane is in general a difficult task.  Indeed, first
of all, even for one brane the gauge group is in general non abelian:
namely it is the product of $SU(2)$, $U(2)$ and $O(2)$ gauge groups, 
and hence the equation defining the moduli space are two by
two matrix equations. Moreover the moduli space of an $\mathcal{N}=1$
field theory in three dimension is real and non-holomorphic and hence
one cannot use the powerful tools of the algebraic complex geometry.
Following \cite{Forcella:2009jj} we proceed as follows.  We provide an
ansatz for the two by two matrices describing the matter fields of the
$\mathcal{N}=1$ theory in terms of the complex scalar fields of the
$\mathcal{N}=2$ parent theory for one M2 brane. It follows that the
zero potential condition for the $\mathcal{N}=1$ theory reproduces
exactly the same equations of the parent theory (\ref{VB0}) in terms
of the ansatz fields. We then verify that the ansatz exhausts the
vacuum space of the $\mathcal{N}=1$ theory, i.e. that there are no
other connected flat directions.

The moduli space is obtained by quotienting by the action of
the gauge group.  The ansatz we use is perfectly suited for
this scope. Indeed, as we will explicitly see in the following examples, our
ansatz breaks the gauge group down to its abelian subgroup: a bunch of
$SO(2)$s plus the discrete non-holomorphic $\Theta$.  The $SO(2)$s
that leave the ansatz invariant act as the $U(1)$s of the parent
theory on the ansatz fields.  Hence the quotient by the $SO(2)$s
exactly reproduces the CY$_4$ $Y$ cone quotiented by the additional
discrete action $\mathbb{Z}_k$ associated to the CS levels.  The
remaining discrete action $\Theta$ is generated by the parity
inversion $\sigma_3 \in O(2)$ and the element $i \sigma_3$ of $SU(2)$
and $U(2)$.  This last action exactly generates the needed
anti-holomorphic involution to obtain the Spin$(7)$ cone as explained in
section \ref{sec:aoi}.

By following this procedure we systematically check that the moduli
spaces for one M2 brane for pairs of theories, claimed to be dual, are
the same and that they coincide with the Spin$(7)$ cone obtained by
the anti-holomorphic involution on the CY$_4$ cone of the associated
parent theories. 

In the near horizon limit the AdS/CFT correspondence provides some
arguments to support the fact that the dual pairs of theories
previously constructed are actually two equivalent UV descriptions of
the same IR strong coupling fixed point, dual to M-theory on AdS$_4$
$\times$ $G_2$ background.

\subsubsection*{Relation with Seiberg-like Duality}

When the orientifold action leaves unitary groups we can sometimes 
argue that the Spin$(7)$ duality is a Seiberg-like duality. In this case we can
think to a type IIB brane setup that is locally $\mathcal{N}=2$, but globally  $\mathcal{N}=1$.
Supersymmetry is broken to $\mathcal{N}=1$ because of the orientifold
on some gauge group or on some bifundamental fields not involved in
the duality.  In this case we can move consecutive $(1,p_i)$ branes
and locally reproduce the same transformation as in (\ref{N2r}).  We
claim that the resulting theory is Seiberg-like dual to the first
theory. Indeed it has been obtained by applying the usual rules for
brane exchange and brane creation.

A first check of the duality is that the $\mathcal{N}=1$ theory
obtained by moving the branes is indeed exactly the theory that we
would have obtained instead projecting the Seiberg-like dual theory of
the parent $\mathcal{N}=2$ theory, closing in this way the circle of
dualities.
%
%
%
%
%
%
%
%
%
%
\section{Examples}
\label{sec:examples}

In this section we study examples of Spin$(7)$ dualities between pairs
of three dimensional gauge theories along the lines explained in the
previous section.  We adopt the following strategy.  First we
introduce the $\mathcal{N} = 1$ conjectured dual pairs and then we
show that these models describe the same IR physics.

We show that two conjectured $\mathcal{N}=1$ dual theories can be
obtained by projecting two $\widetilde{L^{aba}}_{k_i}$ $\mathcal{N}=2$
toric dual models.  These $\mathcal{N}=2$ models are toric quiver
gauge theories associated to CY$_4$ singularities.  By projecting
these dual pairs with the anti-holomorphic involution introduced in
section \ref{sec:aoi} we obtain $\mathcal{N} = 1$ dual pairs that
reproduce the same Spin$(7)$ geometry. These models are Spin$(7)$
dual.

First  we present a very simple example. It is a toy model, where the Spin$(7)$ duality actually
coincides with a parity transformation,
that should however help the comprehension of our strategy. In the second example
we  increase the complexity studying a more intricate example of 
Spin$(7)$ duality.
%
%
%
%
%
%
%
%
\subsection{First example}
\label{es:1}

The first Seiberg-like dual pair that we consider consists of
$\mathcal{N}=1$ CS matter theories with three gauge groups as
presented in figure \ref{fig:L2221}.  The gauge groups are
\begin{equation}
O(2N)_{-2k} \times U(2N)_{2k} \times SP(2N)_{-k}
\end{equation}
and four bifundamental fields $\mathcal{Q}_{1}$,
$\widetilde{\mathcal{Q}_{1}}$, $\mathcal{Q}_{2}$ and
$\widetilde{\mathcal{Q}_{2}}$ transforming under
the gauge groups as
\begin{equation}
\begin{array}{c||ccc}
&O(2N)_{-2k} & U(2N)_{2k} & SP(2N)_{-k}\\
\hline
\mathcal{Q}_{1}& \Box  &\Box &\Box \\
\widetilde{\mathcal{Q}_{1}}&\Box &
\overline \Box
 &\Box \\
\mathcal{Q}_{2}&\Box &\Box &\Box \\
\widetilde{\mathcal{Q}_{2}}&\Box &\overline \Box &\Box 
\end{array}
\end{equation}

The $\mathcal{N}=1$ superpotential is
\bea
\label{eq:spotL222ele}
W &=& - \mathcal{Q}_{1} J \widetilde{\mathcal{Q}_{1}^*}
\mathcal{Q}_{1}^* J \widetilde{\mathcal{Q}_{1}} +
\widetilde{\mathcal{Q}_{1}^*} \mathcal{Q}_{2}^*
\widetilde{\mathcal{Q}_{2}^*} \mathcal{Q}_{1}^* - \mathcal{Q}_{2}^*
\widetilde{\mathcal{Q}_{2}} \mathcal{Q}_{2}
\widetilde{\mathcal{Q}_{2}^*} - \widetilde{\mathcal{Q}_{2}}
\mathcal{Q}_{1} \widetilde{\mathcal{Q}_{1}} \mathcal{Q}_{2}
\nonumber\\
&-&\frac{k}{\pi} (R_{SP}^2+R_{O}^2-R_{U}^2)+ R_{O} \left(
  \mathcal{Q}_1^{\dagger} \mathcal{Q}_1 + \mathcal{Q}_1^T
  \mathcal{Q}_1^* - \widetilde{\mathcal{Q}_1}
  \widetilde{\mathcal{Q}_1^{\dagger}} -\widetilde{\mathcal{Q}_1^*}
  \widetilde{\mathcal{Q}_1^T} \right)
\nonumber\\
&+& R_{U} \left( \widetilde{\mathcal{Q}_1^{\dagger}}
  \widetilde{\mathcal{Q}_1} - \mathcal{Q}_1 \mathcal{Q}_1^{\dagger}
  -\widetilde{\mathcal{Q}_2^{\dagger}} \widetilde{\mathcal{Q}_2} +
  \mathcal{Q}_2 \mathcal{Q}_2^{\dagger} \right)
\nonumber\\
&+& R_{SP} \left( \widetilde{\mathcal{Q}_2}
  \widetilde{\mathcal{Q}_2^{\dagger}} -J \widetilde{\mathcal{Q}_2^*}
  \widetilde{\mathcal{Q}_2^{T}} J - \mathcal{Q}_2^\dagger
  \mathcal{Q}_2 +J \mathcal{Q}_2^T \mathcal{Q}_2^* J \right) 
\eea
We claim that this model is Spin$(7)$ dual to another
$\mathcal{N}=1$ CS matter theory with gauge groups:
\begin{equation}
O(2N)_{2k} \times U(2N)_{-2k} \times SP(2N)_{k}
\end{equation}
with four bifundamental fields ${\mathcal{Q}_{d}}_{i},\widetilde
{\mathcal{Q}_{d}}_{i}$, $i=1,2$, as in figure \ref{fig:L2221}, and the
$\mathcal{N}=1$ dual superpotential coincides with
(\ref{eq:spotL222ele}) with $k \rightarrow -k$.  These two models can
be obtained by projecting two toric dual parent $\mathcal{N}=2$
theories.

\subsubsection*{$\mathcal{N }=2$ parents}

The parent $\mathcal{N}=2$ theories are denoted as
$\widetilde{L^{222}_{k_i}}$ theory.  There are two possible quivers
associated to this singularity, each with four gauge groups. One has
eight bifundamentals and quartic couplings and the second one has
eight bifundamentals and two adjoints. Here we analyze the moduli
space for one M2 brane, where the gauge group is simply $U(1)^4$. The
moduli space for the $U(N)$ case is the $N$-times symmetric product
of the moduli space for a single brane.

At this point of the discussion we specify a choice of CS levels
 useful to perform the orientifold.  We choose the levels as
$\vec k=(k,-k,k,-k)$. 
 \bef
\includegraphics[width=15cm]{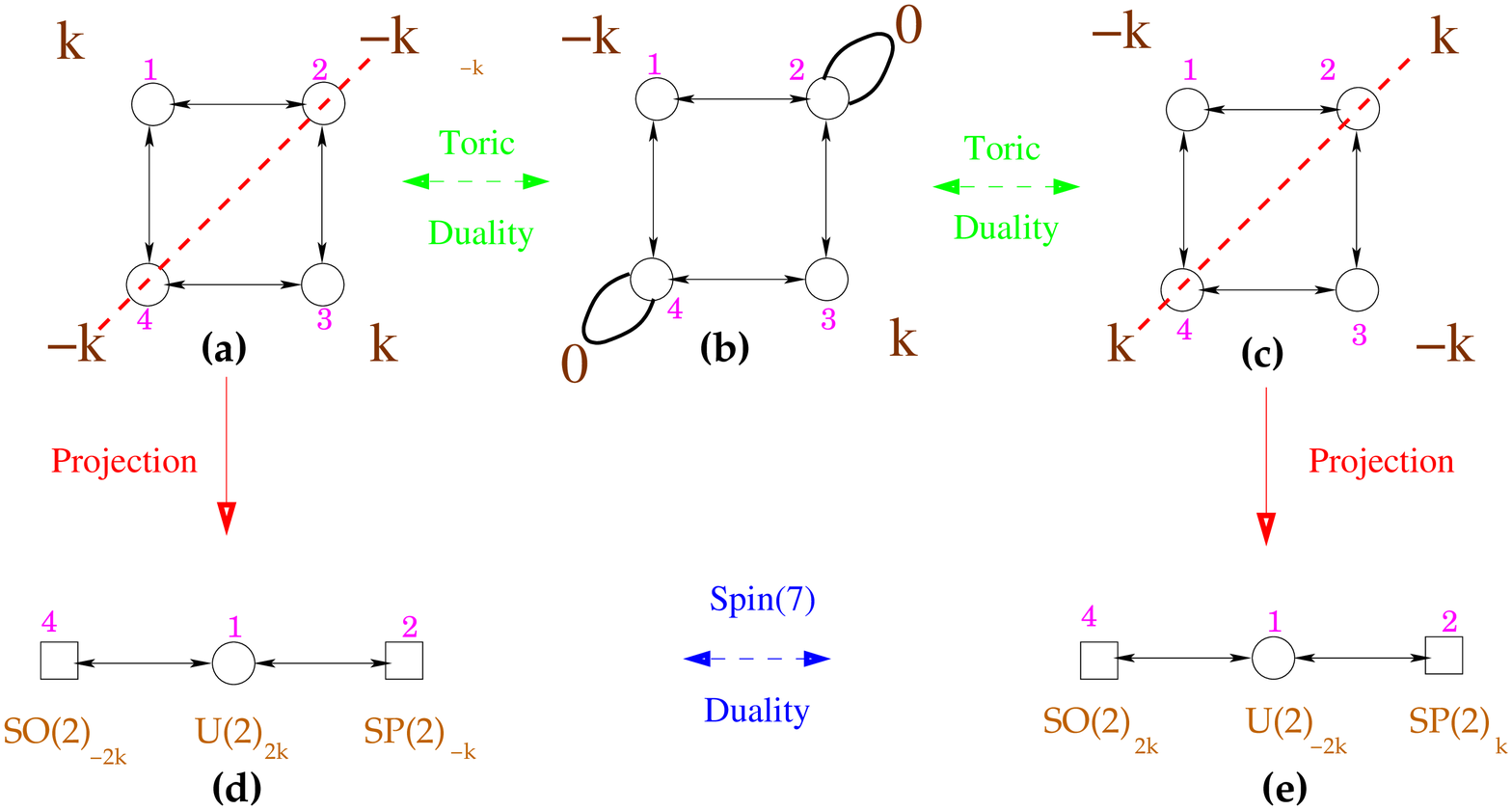}
\caption{ This picture represents schematically the relation between
  toric duality and Spin$(7)$ duality.  The models in $(a)$, $(b)$ and
  $(c)$ represent three toric dual phases.  In the cases $(a)$ and
  $(c)$ the orientifold projection acts by folding the quiver, along
  the dashed red lines. By projecting these models to $\mathcal{N}=1$
  Spin$(7)$ cones we obtain the phases $(d)$ and $(e)$, that are
  related by Spin$(7)$ duality.}
\label{fig:L2221}
\eef
The $\mathcal{N}=2$ superpotential for the first phase is 
\be 
\label{WI}
W_{I} = Q_{12} Q_{23} Q_{32} Q_{21} -Q_{23} Q_{34} Q_{43} Q_{32} +
Q_{34}Q_{41}Q_{14}Q_{43}-Q_{41}Q_{12}Q_{21}Q_{14} 
\ee
The equations of motion are solved by
\begin{equation}
Q_{12} Q_{21} = Q_{34} Q_{43} 
\quad,
\quad
Q_{23} Q_{32}  = Q_{14}Q_{41}
\end{equation}
The operators gauge invariant with respect to the gauge factors  orthogonal to the CS vector
are
\bea
x_1 &=& Q_{12} Q_{21} \quad
x_2 = Q_{23} Q_{32} \quad
x_3 = Q_{34} Q_{43} \quad
x_4 = Q_{14} Q_{41} \non \\
x_5 &=& Q_{12} Q_{34} \quad
x_6 = Q_{21} Q_{43} \quad
x_7 = Q_{23} Q_{41} \quad
x_8 = Q_{32} Q_{14}
\eea
They are related by
\begin{equation}
\label{eq:geoCY4}
x_1 x_3 = x_1^2 = x_5 x_6 \quad\quad\quad
x_2 x_4 = x_2^2 = x_7 x_8
\end{equation}
These equations define the CY$_4$ $Y$ that has to be mod by the $\mathbb{Z}_k$
along the direction of the CS.

The second quiver is represented in figure \ref{fig:L2221} (b).  It
has superpotential
\be W_{II} =
Q_{12} \Phi_2 Q_{21} -Q_{32} \Phi_2 Q_{23}+Q_{23} Q_{34} Q_{43} Q_{32}
+ Q_{34} \Phi_4 Q_{43}-Q_{14} \Phi_4 Q_{41}+Q_{41}Q_{12}Q_{21}Q_{14}
\ee
The $U(1)$ gauge groups have CS levels $\vec k = (-k,0,k,0)$.
One can check that this model describes the same CY$_4$ geometry
(\ref{eq:geoCY4}) of the original theory, and the two phases are toric
dual.
One can build another dual phase with superpotential (\ref{WI}). The
$U(1)$ gauge groups have CS levels $\vec k = (-k,k,-k,k)$. The quiver
is represented in \ref{fig:L2221} (c).  This is the other parent
theory that we have to project to obtain the Spin$(7)$ dual phase.

The next step consists of studying the orientifold projection of the
dual models represented in figure \ref{fig:L2221} (a) and (c) to
$\mathcal{N}=1$ and check the Spin$(7)$ duality between the models 
represented in  figure \ref{fig:L2221} (d) and (e).

\subsubsection*{Projection to $\mathcal{N}=1$.}

We start by analyzing the first case.
The anti-holomorphic involution on the coordinates is
\begin{eqnarray}
\label{act1S}
&&
x_1 \rightarrow -  x_2^* \quad
x_2 \rightarrow ~~ x_1^* \quad
x_3 \rightarrow - x_4^* \quad
x_4 \rightarrow ~~ x_3^*
\nonumber\\
&&
x_5 \rightarrow ~~ x_8^* \quad
x_6 \rightarrow ~~ x_7^* \quad
x_7 \rightarrow - x_6^* \quad
x_8 \rightarrow - x_5^*
\end{eqnarray}
 this action does not have fixed points, except the origin, on the CY$_4$ geometry (\ref{eq:geoCY4}).

The anti-holomorphic involution acts on the gauge fields and on the
scalars $\sigma_i$ as
\begin{equation}
\begin{array}{cccccc}
  A_{\mu}^1   & \rightarrow &-\Omega (A_\mu^{3})^T \Omega^{-1} &
  \quad\quad\quad A_{\mu}^{2,4} &\rightarrow &-\Omega_{2,4} (A_\mu^{2,4})^T \Omega_{2,4}^{-1}  \\
  \sigma_1   & \rightarrow& \Omega \sigma_3^T \Omega^{-1} &
  \quad\quad\quad \sigma_{2,4} &\rightarrow &\Omega_{2,4} \sigma_{2,4}^T \Omega_{2,4}^{-1}  \\
\end{array}
\end{equation}
The anti-holomorphic involution acts on the bifundamental as 
\be
\begin{array}{llll}
Q_{41} \rightarrow  ~~ \Omega_4   Q_{43}^*  \Omega^{-1}\quad  &
Q_{14} \rightarrow  ~~ \Omega     Q_{34}^*  \Omega_4^{-1}\quad & 
Q_{43} \rightarrow  ~~ \Omega_4   Q_{41}^*  \Omega^{-1}\quad &
Q_{34} \rightarrow - \Omega     Q_{14}^*  \Omega_4^{-1}
 \\
Q_{23} \rightarrow - \Omega_2  Q_{21}^* \Omega^{-1} \quad &
Q_{32} \rightarrow - \Omega    Q_{12}^* \Omega_2^{-1} \quad &
Q_{12} \rightarrow - \Omega    Q_{32}^* \Omega_2^{-1}\quad  &
Q_{21} \rightarrow  ~~ \Omega_2  Q_{23}^* \Omega^{-1}  
\end{array}
\ee 
%

The transformation corresponds to an orientifold projection, sending
$\sigma \rightarrow \sigma$ and the angle $\psi \rightarrow -\psi$.
Moreover this transformation is a symmetry of the $\mathcal{N}=2$
lagrangian.  Indeed one can check that the superpotential is sent into
its complex conjugate and that the $D$-terms transform consistently
with the constraint $D_a=\frac{k_a}{2 \pi} \sigma_a$. From now on we
fix $\Omega_4=I_{2N}$ and $\Omega_2=J_{2N}$, and the projected
gauge theory is $ O(2N)_{-2k} \times U(2N)_{2k} \times SP(2N)_{-k} $.
This is precisely the gauge symmetry of the $\mathcal{N}=1$ theory
that we want to obtain.

The dual phase can be obtained analogously. The projection is obtained
by flipping the sign of each CS level. This does not affect the ansatz
but only the $D$ terms.

\subsubsection*{Calculation of the $\mathcal{N}=1$ moduli space}

The next step consists of calculating the moduli space for a single M2
brane and show that they coincide. In this case even for the single
brane the gauge group is non abelian, $ O(2)_{-2k} \times U(2)_{2k}
\times SP(2)_{-k} $.

We choose an ansatz for the $\mathcal{N}=1$ fields
$\mathcal{Q}_{1}$, $\widetilde{\mathcal{Q}_{1}}$, $\mathcal{Q}_{2}$
and $\widetilde{\mathcal{Q}_{2}}$ in terms of the $\mathcal{N}=2$
bifundamentals as
\begin{eqnarray}
\label{ansatz1}
\mathcal{Q}_{1} = \frac{Q_{12}+Q_{32}^*}{2} I +\frac{Q_{12}-Q_{32}^*}{2i} J
&\quad \quad&
\widetilde {\mathcal{Q}_{1}} = \frac{Q_{21}+Q_{23}^*}{2} I +\frac{Q_{21}-Q_{23}^*}{2i} J
\nonumber\\
\widetilde { \mathcal{Q}_{2}} = \frac{Q_{41}+i Q_{43}^*}{2} I +\frac{Q_{41}-i Q_{43}^*}{2i} J
&\quad \quad&
\mathcal{Q}_{2} = \frac{Q_{14}- i Q_{34}^*}{2} I +\frac{Q_{14}+Q_{34}^*}{2i} J
\nonumber\\
\end{eqnarray}
The ansatz (\ref{ansatz1}) exactly reproduces the equations of motion
(\ref{VB0}) of the parent $\mathcal{N}=2$ theory.
Moreover this ansatz exhausts the vacuum space of the
$\mathcal{N}=1$ theory.  Indeed we checked that by fluctuating around
the solution there are not other flat directions.

There are four residual abelian gauge factors on the moduli space,
$SO(2) \in O(2)$, $SO(2) \in SP(2)$ and $U(1)^2 \in U(2)$. They act on the ansatz fields 
exactly as the $U(1)^4$ 
gauge group in the $\mathcal{N}=2$ case. One can observe that one of them acts
trivially, two combinations are used to mod the moduli space and the
last factor is broken to $\mathbb{Z}_{2k}$ by the CS. 
In this way the moduli space exactly reproduces the geometry (\ref{eq:geoCY4}) modded by $\mathbb{Z}_{2k}$ as in the 
parent theory.  

However there is still a residual discrete symmetry, $\Theta$, generated by
$\sigma_3$ in $O(2)$ and $i \sigma_3$ in $SU(2)$ and $U(2)$. It
corresponds to the antiholomorphic involution.
The moduli space of the $\mathcal{N}=1$ theory is then the Spin$(7)$ quotient geometry $Y/\Theta_k$, 
where $\Theta_k$ is the combination of $\Theta$ in (\ref{act1S}) with $\mathbb{Z}_{2k}$, and Y is the CY$_4$ in 
 (\ref{eq:geoCY4}).

The analysis of the dual phase is similar.  The moduli space of the
two $\mathcal{N}=1$ theories are coincident and this supports the
Spin$(7)$ duality for this first simple example.  Observe that the
duality can be understood as a parity transformation $k \rightarrow
-k$.  Although the simplicity of this duality, we studied this toy
model because we believe that it could be useful for the reader 
to understand our general picture.
%
%
%
%
%
%
%
%
%
%
\subsection{Second example}
\label{es:3}

Let us now provide a more involved and interesting example of Spin$(7)$ duality.
In this case we consider $\mathcal{N}=1$
models with only orthogonal or symplectic gauge groups.  We consider a
case with four gauge groups.  We distinguish two possibilities
\begin{eqnarray}
\left\{
\begin{array}{cl}
(I)&O(2N)_{2k} \times O(2N)_{0} \times O(2N)_{-2k} \times O(2N)_{0} 
\\
(II)&SP(2N)_{k} \times SP(2N)_{0} \times SP(2N)_{-k} \times SP(2N)_{0}
\end{array}
\right.
\end{eqnarray}
In the rest of the section we study the case $I$ but everything can be
easily generalized to the case $II$.  The $\mathcal{N}=1$ superpotential is\footnote{With abuse of notation we keep the same notation as before for the the matter fields $\mathcal{Q}_{ij}$ even if both indices $i$ and $j$ refer now to the fundamental representation because the gauge group is now real.}
\begin{eqnarray}
\label{es3ele}
W &=& \mathcal{Q}_{12}\mathcal{Q}_{23}\mathcal{Q}_{32}\mathcal{Q}_{21}-\mathcal{Q}_{23}\mathcal{Q}_{34}\mathcal{Q}_{43}\mathcal{Q}_{32} + \mathcal{Q}_{34}\mathcal{Q}_{41}\mathcal{Q}_{14}\mathcal{Q}_{43}-\mathcal{Q}_{41}\mathcal{Q}_{12}\mathcal{Q}_{21}\mathcal{Q}_{14}
\nonumber\\
&+&R_1( \mathcal{Q}_{12} \mathcal{Q}_{12}^T - \mathcal{Q}_{21}^T \mathcal{Q}_{21} +\mathcal{Q}_{14} \mathcal{Q}_{14}^T-\mathcal{Q}_{41}^T \mathcal{Q}_{41})
\nonumber\\
&+&R_2 (\mathcal{Q}_{21}\mathcal{Q}_{21}^T -\mathcal{Q}_{12}^T\mathcal{Q}_{12}+\mathcal{Q}_{23} \mathcal{Q}_{23}^T-\mathcal{Q}_{32}^T\mathcal{Q}_{32})
\nonumber\\
&+&
R_3(\mathcal{Q}_{32} \mathcal{Q}_{32}^T-\mathcal{Q}_{23}^T \mathcal{Q}_{23}+\mathcal{Q}_{34} \mathcal{Q}_{34}^T-\mathcal{Q}_{43}^T \mathcal{Q}_{43})
\nonumber\\
&+&
R_4 (\mathcal{Q}_{43} \mathcal{Q}_{43}^T+ \mathcal{Q}_{34}^T \mathcal{Q}_{34}-\mathcal{Q}_{41} \mathcal{Q}_{41}^T+\mathcal{Q}_{14}^T \mathcal{Q}_{14})   
\nonumber\\
&+&\frac{k}{2\pi}(R_1^2-R_3^2)
\end{eqnarray}
The Spin$(7)$ dual theories have gauge groups
\begin{eqnarray}
\left\{
\begin{array}{cl}
(I)&O(2N)_{-2k} \times O(2N)_{2k} \times O(2N)_{-2k} \times O(2N)_{2k} 
\\
(II)&SP(2N)_{-k} \times SP(2N)_{k} \times SP(2N)_{-k} \times SP(2N)_{k}
\end{array}
\right.
\end{eqnarray}
Here we still restrict to the first case.  The $\mathcal{N}=1$ superpotential becomes
\begin{eqnarray}\label{es3mag}
  W &=&\mathcal{Q}_{14}\mathcal{Q}_{44}\mathcal{Q}_{41} -\mathcal{Q}_{12}\mathcal{Q}_{22}\mathcal{Q}_{21} + \mathcal{Q}_{32}\mathcal{Q}_{22}\mathcal{Q}_{23}-\mathcal{Q}_{23}\mathcal{Q}_{34}\mathcal{Q}_{43}\mathcal{Q}_{32} + \mathcal{Q}_{34}\mathcal{Q}_{44}\mathcal{Q}_{43}
  \nonumber\\
  &+&R_1( \mathcal{Q}_{12} \mathcal{Q}_{12}^T - \mathcal{Q}_{21}^T \mathcal{Q}_{21} +\mathcal{Q}_{14} \mathcal{Q}_{14}^T-\mathcal{Q}_{41}^T \mathcal{Q}_{41})
  \nonumber\\
  &+&R_2 (\mathcal{Q}_{21}\mathcal{Q}_{21}^T -\mathcal{Q}_{12}^T\mathcal{Q}_{12}+\mathcal{Q}_{23} \mathcal{Q}_{23}^T-\mathcal{Q}_{32}^T\mathcal{Q}_{32})
  \nonumber\\
  &+&
  R_3(\mathcal{Q}_{32} \mathcal{Q}_{32}^T-\mathcal{Q}_{23}^T \mathcal{Q}_{23}+\mathcal{Q}_{34} \mathcal{Q}_{34}^T-\mathcal{Q}_{43}^T \mathcal{Q}_{43})
  \nonumber\\
  &+&
  R_4 (\mathcal{Q}_{43} \mathcal{Q}_{43}^T+ \mathcal{Q}_{34}^T \mathcal{Q}_{34}-\mathcal{Q}_{41} \mathcal{Q}_{41}^T+\mathcal{Q}_{14}^T \mathcal{Q}_{14})   
  \nonumber\\
  &+&\frac{k}{2\pi}(R_4^2-R_1^2+R_2^2-R_3^2)
\end{eqnarray}
In the rest of this section we study this duality as before.  First we
provide the $\mathcal{N}=2$ dual parents, then we study the projection
to $\mathcal{N}=1$ and show that the moduli spaces match, supporting the
Spin$(7)$ duality.

\subsubsection*{$\mathcal{N}=2$ parents}

In this case the parent theories are $\widetilde{L^{222}_{k_i}}$
models in the $\mathcal{N}=2$ case. The dual phase is obtained by
dualizing the first gauge group. 
The quiver and the superpotential coincide with the ones studied
in subsection \ref{es:1}.

We study here the moduli space for one M2 brane where the 
gauge group is  $U(1)^4$ gauge group.
The CS levels are $\vec k = (k,0,-k,0)$.  The gauge invariant
combinations, orthogonal to the CS vector, are
\begin{eqnarray}
x_1 &=& Q_{12} Q_{21} = Q_{34} Q_{43}\quad \quad \quad \quad \quad
x_2 = Q_{23} Q_{32} = Q_{14} Q_{41} \nonumber \\
y_1 &=& Q_{12} Q_{23} \quad \quad
y_2 = Q_{21} Q_{32} \quad \quad
y_3 = Q_{34} Q_{41} \quad \quad
y_4 = Q_{43} Q_{14} \nonumber \\
\end{eqnarray}
They are related by
\begin{equation}
\label{cy42}
x_1 x_2 = y_1 y_2 = y_3 y_4
\end{equation}
These equations define the CY$_4$ $Y$ that has to be mod by the $\mathbb{Z}_k$.

The $U(1)$ gauge groups of the toric dual $\mathcal{N}=2$ phase
have CS levels $\vec k = (-k,k,-k,k)$.
The gauge invariant combinations, orthogonal to the CS vector, are
\begin{eqnarray}
x_1 &=& Q_{12} Q_{21} =  Q_{23} Q_{32} = Q_{44}  \quad \quad 
x_2  =  Q_{34} Q_{43} =  Q_{14} Q_{41}  =Q_{22} \nonumber \\
y_1 &=& Q_{12} Q_{34}   \quad \quad
y_2  =  Q_{21} Q_{43}  \quad  \quad 
y_3  =  Q_{23} Q_{41}   \quad  \quad
y_4  =  Q_{32} Q_{14} \nonumber \\
\end{eqnarray}
They are related by
\begin{equation}
\label{cy2}
x_1 x_2 = y_1 y_2 = y_3 y_4
\end{equation}
These equations define the CY$_4$ $Y$ that has to be mod by the
$\mathbb{Z}_k$.  The moduli space of the two theories is then the same
and they are indeed toric dual.

In the rest of this section we project the theories to $\mathcal{N}=1$
to obtain the two models discussed above. We check that they reproduce
the expected $\mathcal{N}=1$ phases and compute the classical moduli
space with our usual procedure. Eventually we match the two moduli
spaces, supporting the Spin$(7)$ duality.

\subsubsection*{Projection to $\mathcal{N}=1$  of the electric phase}

We choose the anti-holomorphic involution as
\begin{eqnarray}
\label{act22}
x_1 \rightarrow -x_1^*\quad,\quad 
x_2 \rightarrow -x_2^*\quad&,&\quad 
x_3 \rightarrow -x_3^*\quad,\quad 
x_4 \rightarrow -x_4^*
\nonumber \\
y_1 \rightarrow -y_1^*\quad,\quad 
y_2 \rightarrow -y_2^*\quad&,&\quad 
y_3 \rightarrow -y_3^*\quad,\quad 
y_4 \rightarrow -y_4^*
\end{eqnarray}
%
this action has a real 
four dimensional locus of fixed points on the CY$_4$ geometry (\ref{cy2}).

On the fields $Q_{ij}$ this anti-involution becomes
\begin{equation}
\begin{array}{rccl}
Q_{12} \rightarrow -\Omega_{1} Q_{12}^{*} \Omega_{2}^{-1}&
Q_{21} \rightarrow ~\Omega_{2} Q_{21}^{*} \Omega_{1}^{-1}&
Q_{23} \rightarrow ~\Omega_{2} Q_{23}^{*} \Omega_{3}^{-1}&
Q_{32} \rightarrow -\Omega_{3} Q_{32}^{*} \Omega_{2}^{-1}\\
Q_{34} \rightarrow -\Omega_{3} Q_{34}^{*} \Omega_{4}^{-1}&
Q_{43} \rightarrow ~\Omega_{4} Q_{43}^{*} \Omega_{3}^{-1}&
Q_{41} \rightarrow ~\Omega_{4} Q_{41}^{*} \Omega_{1}^{-1}&
Q_{14} \rightarrow -\Omega_{1} Q_{14}^{*} \Omega_{4}^{-1}
\end{array}
\end{equation}
Here $\Omega_i=I_2$ or $J_2$ means that we project on an orthogonal or
symplectic group.  By choosing $\Omega_{i}=I_{2N}$ the gauge groups become 
$O(2N)$ with $\vec k=(2k,0,-2k,0)$ while choosing $\Omega_{i}=I_{2N}$
we have a product of  $SP(2N)$ gauge groups with $\vec k=(k,0,-k,0)$ . 
Also in this case the anti-holomorphic
action is a symmetry of the full lagrangian.
The transformation corresponds to an orientifold projection, sending
$\sigma \rightarrow \sigma$ and the angle $\psi \rightarrow -\psi$.

\subsubsection*{Moduli space of the $\mathcal{N}=1$  electric phase}

Here we compute the moduli space for a single M2 brane.
We choose the ansatz for the $\mathcal{N}=1$ fields as
\begin{equation}
\label{AN}
\mathcal{Q}_{ij} = Re(Q_{ij}) I + Im(Q_{ij}) J 
\end{equation}
Once we plug these projection in the superpotential (\ref{es3ele}) they
reproduce the equations of motion (\ref{VB0}) of the $\mathcal{N}=2$ case.
Moreover this ansatz exhausts the vacuum space of the $\mathcal{N}=1$ theory.

There are four residual abelian $SO(2)$ gauge factors on the moduli
space that act as the $U(1)$ gauge groups in the $\mathcal{N}=2$ case. One of them acts
trivially, two combinations are used to mod the moduli space and the
last factor is broken to $\mathbb{Z}_{2k}$ by the CS.  There is still
a residual discrete symmetry, $\Theta$, generated by $\sigma_3$ in
$O(2)$ that corresponds to the antiholomorphic involution.  The moduli
space is the Spin$(7)$ quotient $Y/\Theta_k$, where $\Theta_k$ is the
combination of the $\Theta$ action (\ref{act22}) with $\mathbb{Z}_{2k}$, and $Y$ is the 
CY$_4$ in (\ref{cy42}).

\subsubsection*{The magnetic  phase}

In this case we choose the anti-holomorphic involution as (\ref{act22}).
On the fields $Q_{ij}$ this anti-involution becomes
\begin{equation}
\begin{array}{cccc}
Q_{12} \rightarrow -\Omega_{1} Q_{12}^{*} \Omega_{2}^{-1}&
Q_{21} \rightarrow \Omega_{2} Q_{21}^{*} \Omega_{1}^{-1}&
Q_{23} \rightarrow \Omega_{2} Q_{23}^{*} \Omega_{3}^{-1}&
Q_{32} \rightarrow -\Omega_{3} Q_{32}^{*} \Omega_{2}^{-1}\\
Q_{34} \rightarrow -\Omega_{3} Q_{34}^{*} \Omega_{4}^{-1}&
Q_{43} \rightarrow \Omega_{4} Q_{43}^{*} \Omega_{3}^{-1}&
Q_{41} \rightarrow \Omega_{4} Q_{41}^{*} \Omega_{1}^{-1}&
Q_{14} \rightarrow -\Omega_{1} Q_{14}^{*} \Omega_{4}^{-1}\\
&
Q_{22} \rightarrow -\Omega_{2} Q_{22}^{*} \Omega_{2}^{-1}&
Q_{44} \rightarrow -\Omega_{4} Q_{44}^{*} \Omega_{4}^{-1}&
\end{array}
\end{equation}
where $\Omega_i$ and the ansatz for the bifundamentals are chosen as
before.  Also the adjoints become $\mathcal{Q}_{ii}= Re(Q_{11}) I_{2} +
Im(Q_{ii}) J_{2}$ and they do not contribute to the $D$-terms.  The ansatz
reproduces the equations of motion (\ref{VB0}) of the $\mathcal{N}=2$
case and it exhausts the vacuum space of the $\mathcal{N}=1$ theory.

There are four residual abelian $SO(2)$ gauge factors on the moduli
space that act as in the $\mathcal{N}=2$ case. One of them acts
trivially, two combinations are used to mod the moduli space and the
last factor is broken to $\mathbb{Z}_{2k}$ by the CS.  There is still
a residual discrete symmetry, $\Theta$, generated by $\sigma_3$ in
$O(2)$ that corresponds to the antiholomorphic involution. 
As in the electric phase the moduli space for the magnetic phase is the Spin$(7)$ quotient 
$Y/\Theta_k$, where $\Theta_k$ is the
combination of the $\Theta$ action (\ref{act22}) with $\mathbb{Z}_{2k}$, and $Y$ is the 
CY$_4$ in (\ref{cy42}).

The two geometries coincide and this confirms that the two
$\mathcal{N}=1$ theories are Spin$(7)$ dual.

As already remarked the Spin$(7)$ duality is insensitive to the
presence of fractional branes.  The choice of equal rank, $2N$, for
each gauge factor in the examples studied above comes naturally from
the orientifold projection. However the Spin$(7)$ duality would have
been valid also for different choices of ranks for the projected
theories. In the next section we explore the possibility to fix the
ranks, and hence the number of fraction branes, using Seiberg-like
dualities.
%
%
%
%
%
%
%
%
%
%
%
%
%
%
%
\section{Spin $(7)$ duality as Seiberg like duality}
\label{sec:geno}

For $\mathcal{N}=2$ CS-matter theories in \cite{Amariti:2009rb} it has
been shown that some toric dualities between $\widetilde{L_{k}^{aba}}$
theories are actually three dimensional Seiberg-like dualities. Namely
that the two different field theories not only have the same moduli
space, but they are actually two different descriptions of the same IR
conformal field theory that is holographic dual to the M theory
background: AdS$_4$ $\times$ $H_7$.  It is maybe worth to underline
the principal differences between toric (and similarly Spin$(7)$)
duality and Seiberg-like duality.  Toric duality is essentially the
statement that the moduli space for one regular brane is the same for
the dual pairs of theories.  Seiberg-like duality is instead a
non-abelian statement valid for the set of regular and fractional
branes at the tip of a CY$_4$ or Spin$(7)$ cone.  Indeed, as we have
previously explained, for $\mathcal{N}=2$ $\widetilde{L_{k_i}^{aba}}$
theories the Seiberg-duality transforms the gauge groups as in
(\ref{N2r}).  Anyway, in toric duality, the extra shift in the rank of
the dual gauge group does not play any role.  Indeed the moduli space
for one M2 brane is obtained by setting $N=1$ in all the gauge group
factors and disregarding the rank difference among the various gauge
group factors: only regular M2 branes can explore the geometry
transverse to the brane, while the fractional branes are stacked at
the singularity and do not contribute to the moduli space. Moreover
for $\mathcal{N}=2$ the moduli space of N regular branes is simply the
N times symmetric product of the moduli space for one brane.

In analogy with the $\mathcal{N}=2$ case, in this section we study
examples of Spin$(7)$ dual $\mathcal{N}=1$ pairs of theories that are
also Seiberg-like dual.

\subsection{Example}
\label{es:2}

Let us illustrate in detail a specific example to explain our general
philosophy.  We consider a three dimensional $\mathcal{N}=1$ CS-matter
theory with four gauge groups
\be 
U(2N)_{2k} \times U(2N)_{-2k} \times U(2N)_{0} \times U(2N)_0
\ee 
%
and $\mathcal{N}=1$ superpotential: 
\bea
\label{WIA}
W&=&
\mathcal{Q}_{11} \mathcal{Q}_{12} \mathcal{Q}_{21} \mathcal{Q}_{11}^*
-
\mathcal{Q}_{12} \mathcal{Q}_{23} \mathcal{Q}_{32} \mathcal{Q}_{21} 
+
\mathcal{Q}_{23} \mathcal{Q}_{34} \mathcal{Q}_{43} \mathcal{Q}_{32} 
-
\mathcal{Q}_{34} \mathcal{Q}_{44}^* \mathcal{Q}_{44} \mathcal{Q}_{43} 
\nonumber \\
&+&
\mathcal{Q}_{11}^*  \mathcal{Q}_{12}^*  \mathcal{Q}_{21}^*  \mathcal{Q}_{11}
-
\mathcal{Q}_{12}^*  \mathcal{Q}_{23}^*  \mathcal{Q}_{32}^*  \mathcal{Q}_{21}^* 
+
\mathcal{Q}_{23}^*  \mathcal{Q}_{34}^*  \mathcal{Q}_{43}^*  \mathcal{Q}_{32}^* 
-
\mathcal{Q}_{34}^*  \mathcal{Q}_{44}  \mathcal{Q}_{44}^* \mathcal{Q}_{43}^*  
\nonumber \\
&+&
R_1( 
 \mathcal{Q}_{12} \mathcal{Q}_{12}^\dagger - \mathcal{Q}_{21}^\dagger \mathcal{Q}_{21} 
[\mathcal{Q}_{11}, \mathcal{Q}_{11}^\dagger])
+
R_2 (\mathcal{Q}_{21} \mathcal{Q}_{21}^\dagger - \mathcal{Q}_{12}^\dagger \mathcal{Q}_{12} 
              + \mathcal{Q}_{23} \mathcal{Q}_{23}^\dagger - \mathcal{Q}_{32}^\dagger \mathcal{Q}_{32})
              \nonumber \\
&+&
R_3 (\mathcal{Q}_{23} \mathcal{Q}_{23}^\dagger - \mathcal{Q}_{32}^\dagger \mathcal{Q}_{32} 
              + \mathcal{Q}_{43} \mathcal{Q}_{43}^\dagger - \mathcal{Q}_{34}^\dagger \mathcal{Q}_{34})
  +
R_4  
 (\mathcal{Q}_{34} \mathcal{Q}_{34}^\dagger - \mathcal{Q}_{43}^\dagger \mathcal{Q}_{43} 
+[\mathcal{Q}_{44}, \mathcal{Q}_{44}^\dagger]
\nonumber \\
&+&\frac{k}{2\pi} (R_1^2-R_2^2)
\eea
We claim that this theory is Seiberg-like dual to another $\mathcal{N}=1$ CS-matter theory with gauge group and CS levels:
\be 
U(2N)_0 \times U(2(N+|k|))_{2k} \times U(2N)_{-2k} \times U(2N)_0
\ee
and $\mathcal{N}=1$ superpotential
\bea
\label{WIIA}
W &=& 
 \mathcal{Q}_{11}(\mathcal{Q}_{12}\mathcal{Q}_{21}-\mathcal{X}_{11}\mathcal{X}_{11}^*)
-\mathcal{Q}_{12}\mathcal{Q}_{23}\mathcal{Q}_{32}\mathcal{Q}_{21} 
+\mathcal{Q}_{33}(\mathcal{Q}_{32}\mathcal{Q}_{23}-\mathcal{Q}_{34}\mathcal{Q}_{43})
+\mathcal{Q}_{34}\mathcal{X}_{44}\mathcal{X}_{44}^* \mathcal{Q}_{43}
\nonumber \\
&+&
\mathcal{Q}_{11}^*(\mathcal{Q}_{12}^*\mathcal{Q}_{21}^*-\mathcal{X}_{11}^*\mathcal{X}_{11})
-\mathcal{Q}_{12}^*\mathcal{Q}_{23}^*\mathcal{Q}_{32}^*\mathcal{Q}_{21}^* 
+\mathcal{Q}_{33}^*(\mathcal{Q}_{32}^*\mathcal{Q}_{23}^*-\mathcal{Q}_{34}^*\mathcal{Q}_{43}^*)
+\mathcal{Q}_{34}^*\mathcal{X}_{44}^*\mathcal{X}_{44} \mathcal{Q}_{43}^*
\nonumber\\
&+&
R_1 
 (\mathcal{Q}_{12} \mathcal{Q}_{12}^\dagger - \mathcal{Q}_{21}^\dagger \mathcal{Q}_{21} 
-[\mathcal{Q}_{11},\mathcal{Q}_{11}^\dagger])
+
R_2 ( \mathcal{Q}_{21} \mathcal{Q}_{21}^\dagger - \mathcal{Q}_{12}^\dagger \mathcal{Q}_{12} 
              + \mathcal{Q}_{23} \mathcal{Q}_{23}^\dagger - \mathcal{Q}_{32}^\dagger \mathcal{Q}_{32})
\nonumber \\
&+&
R_3 ( \mathcal{Q}_{23} \mathcal{Q}_{23}^\dagger - \mathcal{Q}_{32}^\dagger \mathcal{Q}_{32} 
              + \mathcal{Q}_{43} \mathcal{Q}_{43}^\dagger - \mathcal{Q}_{34}^\dagger \mathcal{Q}_{34})     
+
R_4 (
 \mathcal{Q}_{34} \mathcal{Q}_{34}^\dagger - \mathcal{Q}_{43}^\dagger \mathcal{Q}_{43} 
-[\mathcal{Q}_{44}, \mathcal{Q}_{44}^\dagger]
\nonumber\\
&+&
\frac{k}{2 \pi}(R_2^2-R_3^2)
\eea
We start showing that the two models can be obtained by projecting two
toric dual $\mathcal{N}=2$ parent theories of the
$\widetilde{L_{k}^{444}}$ family.  Then we study the projection and
compute the $\mathcal{N}=1$ moduli space for one M2 brane: namely when
all the gauge groups are $U(2)$.  By comparing the result in the two
phases we show that the two models are indeed Spin$(7)$ dual.
Eventually we show that the brane description supports the claim that
the two models are also Seiberg-like dual.

\subsubsection*{$\mathcal{N}=2$ parents}

The quivers for the parent theories are represented in figure
\ref{fig:L444}. They have eight gauge groups, each associated to a
$U(N_i)_{k_i}$ factor. We choose the ranks as $N_i=N$. In the first
case represented in figure \ref{fig:L444} (a) there is a pair
bifundamental antibifundamental connecting each pair of consecutive
nodes.  The $\mathcal{N}=2$ superpotential is
\bea 
W &=& Q_{12}Q_{23}Q_{32}Q_{21}-Q_{23}Q_{34}Q_{43}Q_{32} 
  + Q_{34}Q_{45}Q_{54}Q_{43}-Q_{45}Q_{56}Q_{65}Q_{54} \nonumber \\
  &+& Q_{56}Q_{67}Q_{76}Q_{65}-Q_{67}Q_{78}Q_{87}Q_{76} 
  + Q_{78}Q_{81}Q_{18}Q_{87}-Q_{81}Q_{12}Q_{21}Q_{18} 
\eea
We choose the CS levels as $\vec k = (-k,k,0,0,0,0,k,-k)$.  Let us
analyze the $\mathcal{N}=2$ moduli space for one M2 regular brane,
namely for the $U(1)^4$ gauge group. After solving the F-term
equations the operators gauge invariant with respect to the gauge
factor orthogonal to the CS vector are
\bea
\label{para}
x_1&=&Q_{12} Q_{21} =  Q_{34} Q_{43} = Q_{56} Q_{65} = Q_{78} Q_{87}\nonumber\\ 
x_2&=&Q_{23} Q_{32} =  Q_{45} Q_{54} = Q_{67} Q_{76} = Q_{81} Q_{18}\nonumber\\
y_1&=&Q_{12} Q_{23} Q_{34} Q_{45} Q_{56} Q_{67} Q_{78} Q_{81}\nonumber\\
y_2&=&Q_{18} Q_{87} Q_{76} Q_{65} Q_{54} Q_{43} Q_{32} Q_{21}\nonumber\\
t_1&=&Q_{12}Q_{87}\quad \quad
t_2=Q_{21}Q_{78} 
\eea
They are related by
\be
\label{eq:geoCY41}
x_1^4 x_2^4 = y_1 y_2 \quad \quad 
t_1 t_2 = x_1^2
\ee
These equations define the CY$_4$ $Y$ that has to be modded by the $\mathbb{Z}_k$ action.

The second parent is obtained by acting with two Seiberg-like
dualities on $U(N_2)$ and $U(N_7)$ respectively.  The dual quiver is
represented in figure \ref{fig:L444} (c).  In this case there are four
extra adjoint fields.  The ranks of the dualized groups are
\begin{equation}
\widetilde{N_2} = N_1+N_3-N_2+|k_2|=N + |k|
\quad,\quad
\widetilde{N_7}=N_6+N_8-N_7+|k_7| = N+|k|
\end{equation}
while all the other ranks remain the same. The CS levels of the dual phase 
are $\vec k = (0,-k,k,0,0,k,-k,0)$.  The
dual $\mathcal{N}=2$ superpotential is
\bea W &=&
Q_{11}(Q_{12}Q_{21}-Q_{18}Q_{81}) -Q_{12}Q_{23}Q_{32}Q_{21}
+Q_{33}(Q_{32}Q_{23}-Q_{34}Q_{43}) +Q_{34}Q_{45}Q_{54}Q_{43}
\nonumber \\
&-& Q_{45}Q_{56}Q_{65}Q_{54} +Q_{66}(Q_{65}Q_{56}-Q_{67}Q_{76})
+Q_{67}Q_{78}Q_{87}Q_{76} -Q_{88}(Q_{87}Q_{78}-Q_{81}Q_{18})
\nonumber\\
\eea
Let us analyze the $\mathcal{N}=2$ moduli space for one M2 regular
brane, namely for the $U(1)^4$ gauge group. Where, as previously
explained, we disregarded the presence of fractional branes, because
they are stacked at the origin and they do not explore the moduli
space.  After solving the F-term equations the gauge invariant
operators orthogonal to the CS vector are
\bea
\label{para2}
 x_1&=&Q_{12}Q_{21} =
Q_{33} = Q_{45}Q_{54} = Q_{66}=Q_{78}Q_{87}=Q_{81}Q_{18}
\nonumber\\
x_2&=&Q_{11}=Q_{23}Q_{32}=Q_{34}Q_{43}=Q_{56}Q_{65}=Q_{67}Q_{76}=Q_{88}
\nonumber\\
y_1&=&Q_{12} Q_{23} Q_{34} Q_{45} Q_{56} Q_{67} Q_{78} Q_{81}\nonumber\\
y_2&=&Q_{18} Q_{87} Q_{76} Q_{65} Q_{54} Q_{43} Q_{32} Q_{21}\nonumber\\
t_1&=&Q_{23}Q_{76} \quad\quad
t_2=Q_{32}Q_{67}
\eea 
and they are related by 
\be 
\label{CYp2}
x_1^4 x_2^4 = y_1 y_2 \quad,\quad t_1 t_2 = x_2^2 \ee These equations
define the CY$_4$ $Y$ that has to be mod by the $\mathbb{Z}_k$.
Equations (\ref{eq:geoCY41}) for the first phase and equations
(\ref{CYp2}) for the second phase are equivalent: the two theories are
indeed Seiberg and toric dual and they have the same moduli space for
one regular brane.
\begin{center}
\bef
\includegraphics[width=16cm]{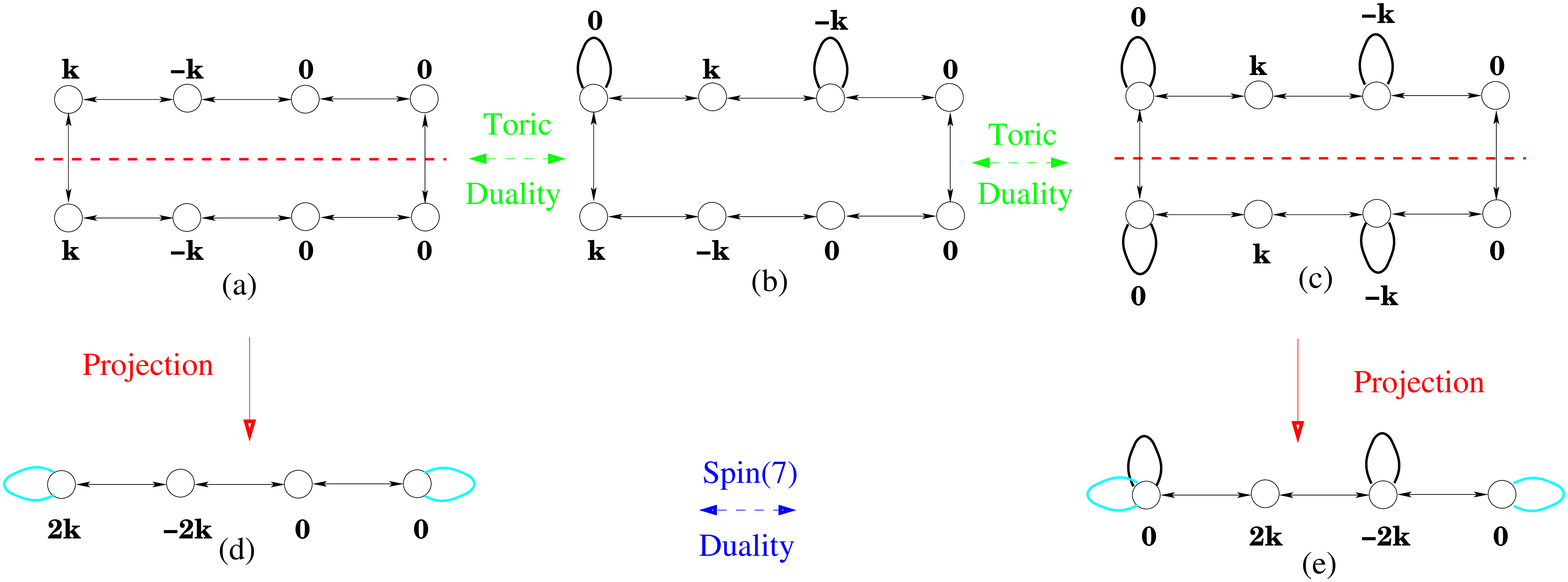}
\caption{ The models in $(a)$, $(b)$ and $(c)$ represent three
  $L^{444}$ toric dual phases.  In the cases $(a)$ and $(c)$ the
  orientifold projection acts by folding the quiver, along the dashed
  red lines. By projecting these models to $\mathcal{N}=1$ Spin$(7)$
  cones we obtain the phases $(d)$ and $(e)$, that are related by
  Spin$(7)$ duality. These models are also Seiberg-like dual.}
\label{fig:L444}
\eef
\end{center}
Here we study the projection of the two phases to obtain the two
$\mathcal{N}=1$ theories introduced above.

\subsubsection*{Projection to $\mathcal{N}=1$ of the electric theory}

In the first case the anti-holomorphic involution on the coordinates is
\be 
\label{actex444}
x_1 \rightarrow x_1^*\quad x_2 \rightarrow -x_2^*\quad y_1
\rightarrow y_2^*\quad y_2 \rightarrow y_1^*\quad t_1 \rightarrow
t_1^*\quad t_2 \rightarrow t_2^* 
\ee
This action has a real four dimensional locus of fixed points on the
CY$_4$ (\ref{eq:geoCY41}) and it represents an orientifold projection that
sends $\sigma \rightarrow \sigma$ and $\psi \rightarrow -\psi$. The
associated orientifold action on the fields is
\bea
&& Q_{12}\rightarrow \phantom{-}\Omega_{1} Q_{87}^* \Omega_{2}^{-1}
\quad Q_{21}\rightarrow \phantom{-}\Omega_{2} Q_{78}^* \Omega_{1}^{-1}
\quad Q_{23}\rightarrow -\Omega_{2} Q_{76}^* \Omega_{3}^{-1} \quad
Q_{32}\rightarrow\phantom{-} \Omega_{3} Q_{67}^* \Omega_{2}^{-1}
\nonumber \\
&& Q_{34}\rightarrow -\Omega_{3} Q_{65}^* \Omega_{4}^{-1}\quad
Q_{43}\rightarrow \phantom{-}\Omega_{4} Q_{56}^* \Omega_{3}^{-1} \quad
Q_{45}\rightarrow \phantom{-}\Omega_{4} Q_{54}^* \Omega_{4}^{-1} \quad
Q_{54}\rightarrow \phantom{-}\Omega_{4} Q_{45}^* \Omega_{4}^{-1}
\nonumber \\
&& Q_{56}\rightarrow\phantom{-} \Omega_{4} Q_{43}^*
\Omega_{3}^{-1}\quad Q_{65}\rightarrow -\Omega_{3} Q_{34}^*
\Omega_{4}^{-1} \quad Q_{67}\rightarrow \phantom{-}\Omega_{3} Q_{32}^*
\Omega_{2}^{-1} \quad Q_{76}\rightarrow - \Omega_{2} Q_{23}^*
\Omega_{3}^{-1}
\nonumber \\
&& Q_{78}\rightarrow \phantom{-}\Omega_{2} Q_{21}^* \Omega_{1}^{-1}
\quad Q_{87}\rightarrow \phantom{-}\Omega_{1} Q_{12}^*
\Omega_{2}^{-1}\quad Q_{81}\rightarrow \phantom{-}\Omega_{1} Q_{18}^*
\Omega_{1}^{-1}\quad Q_{18}\rightarrow \phantom{-}\Omega_{1} Q_{81}^*
\Omega_{1}^{-1}
\nonumber \\
&& Q_{11}\rightarrow -\Omega_{1} Q_{11}^* \Omega_{1}^{-1} \quad
Q_{33}\rightarrow \phantom{-}\Omega_{3} Q_{33}^* \Omega_{3}^{-1} \quad
Q_{66}\rightarrow \phantom{-}\Omega_{6} Q_{66}^* \Omega_{6}^{-1} \quad
Q_{88}\rightarrow -\Omega_{8} Q_{88}^* \Omega_{8}^{-1} \quad
\nonumber\\
\eea 
This action is a symmetry of the $\mathcal{N}=2$
lagrangian.  The superpotential is sent into its complex conjugate and
once again the $D$-terms transform consistently with the constraint
$D_a=\frac{k_a}{2 \pi} \sigma_a$. The gauge groups after the projection
become $U(2N)$ and the CS vector is $\vec k = (-2k,2k,0,0)$.

\subsubsection*{Projection to $\mathcal{N}=1$ of the magnetic theory}

In the dual case the anti-holomorphic involution on the coordinates is
still given by (\ref{actex444}).  On the matter fields is implemented as 
\bea &&
Q_{12}\rightarrow \phantom{-}\Omega_{1} Q_{87}^* \Omega_{2}^{-1} \quad
Q_{21}\rightarrow \phantom{-}\Omega_{2} Q_{78}^* \Omega_{1}^{-1} \quad
Q_{23}\rightarrow -\Omega_{2} Q_{76}^* \Omega_{3}^{-1} \quad
Q_{32}\rightarrow\phantom{-} \Omega_{3} Q_{67}^* \Omega_{2}^{-1}
\nonumber \\
&& Q_{34}\rightarrow -\Omega_{3} Q_{65}^* \Omega_{4}^{-1}\quad
Q_{43}\rightarrow \phantom{-}\Omega_{4} Q_{56}^* \Omega_{3}^{-1} \quad
Q_{45}\rightarrow \phantom{-}\Omega_{4} Q_{54}^* \Omega_{4}^{-1} \quad
Q_{54}\rightarrow \phantom{-}\Omega_{4} Q_{45}^* \Omega_{4}^{-1}
\nonumber \\
&& Q_{56}\rightarrow\phantom{-} \Omega_{4} Q_{43}^*
\Omega_{3}^{-1}\quad Q_{65}\rightarrow -\Omega_{3} Q_{34}^*
\Omega_{4}^{-1} \quad Q_{67}\rightarrow \phantom{-}\Omega_{3} Q_{32}^*
\Omega_{2}^{-1} \quad Q_{76}\rightarrow - \Omega_{2} Q_{23}^*
\Omega_{3}^{-1}
\nonumber \\
&& Q_{78}\rightarrow \phantom{-}\Omega_{2} Q_{21}^* \Omega_{1}^{-1}
\quad Q_{87}\rightarrow \phantom{-}\Omega_{1} Q_{12}^*
\Omega_{2}^{-1}\quad Q_{81}\rightarrow \phantom{-}\Omega_{1} Q_{18}^*
\Omega_{1}^{-1}\quad Q_{18}\rightarrow \phantom{-}\Omega_{1} Q_{81}^*
\Omega_{1}^{-1}
\nonumber \\
&& Q_{11}\rightarrow -\Omega_{1} Q_{11}^* \Omega_{1}^{-1} \quad
Q_{33}\rightarrow \phantom{-}\Omega_{3} Q_{33}^* \Omega_{3}^{-1} \quad
Q_{66}\rightarrow \phantom{-}\Omega_{6} Q_{66}^* \Omega_{6}^{-1} \quad
Q_{88}\rightarrow -\Omega_{8} Q_{88}^* \Omega_{8}^{-1} \quad
\nonumber\\
\eea 
This action as a real 
four dimensional locus of fixed points on the CY$_4$ and it represents an orientifold
projection that sends $\sigma \rightarrow \sigma$ and $\psi
\rightarrow -\psi$.  This action is a symmetry of the $\mathcal{N}=2$
lagrangian.  The superpotential is sent into its complex conjugate and
once again the $D$-terms transform consistently with the constraint
$D_a=\frac{k_a}{2 \pi} \sigma_a$. The gauge groups after the
projection become $U(2N)$ and the CS vector is $\vec k =
(0,-2k,2k,0)$.

\subsubsection*{Moduli space  of the electric $\mathcal{N}=1$ theory}

Here we study the moduli space for a single M2 brane. 
In the projected theory the gauge group is then a product of $U(2)$ factors and the fields are
two by two matrices.
To solve the zero potential condition for the scalar components of
the fields of the $\mathcal{N}=1$ theory we use the ansatz 
%
\bea &&
\mathcal{Q}_{12} =
-\frac{Q_{12}-Q_{87}^*}{2}\sigma_2+\frac{Q_{12}+Q_{87}^*}{2i}\sigma_1
\quad,\quad \mathcal{Q}_{21} =
-\frac{Q_{21}+Q_{78}^*}{2}\sigma_2+\frac{Q_{21}-Q_{78}^*}{2i}\sigma_1
\nonumber \\
&& \mathcal{Q}_{23} = \phantom{-}
\frac{Q_{23}-Q_{76}^*}{2}\sigma_2-\frac{Q_{23}+Q_{76}^*}{2i}\sigma_1
\quad,\quad \mathcal{Q}_{32} = \phantom{-}
\frac{Q_{32}-Q_{67}^*}{2}\sigma_2+\frac{Q_{32}+Q_{67}^*}{2i}\sigma_1
\nonumber \\
&& \mathcal{Q}_{34} = -
\frac{Q_{34}-Q_{65}^*}{2}\sigma_2-\frac{Q_{34}+Q_{65}^*}{2i}\sigma_1
\quad,\quad \mathcal{Q}_{43} = -
\frac{Q_{43}+Q_{56}^*}{2}\sigma_2+\frac{Q_{43}-Q_{56}^*}{2i}\sigma_1
\nonumber \\
\eea
For the other fields we have 
\bea \mathcal{Q}_{11} =
\frac{\sigma_1+i \sigma_2}{2}Q_{81}^* + \frac{\sigma_1-i
  \sigma_2}{2}Q_{18} \quad,\quad \mathcal{Q}_{44} = \frac{\sigma_1-i
  \sigma_2}{2}Q_{45}^* + \frac{\sigma_1+i \sigma_2}{2}Q_{54} 
\eea 
where now the $\sigma_i$ are the two by two Pauli matrices, while the $Q_{ij}$ are 
complex numbers.

By inserting this ansatz in the $\mathcal{N}=1$ superpotential
(\ref{WIA}) we verify that it reproduces the equations of motion (\ref{VB0}) of the
parent $\mathcal{N}=2$ theory.
Moreover we explicitly verified that this ansatz exhausts the vacuum space condition.

To complete the analysis of the moduli space for the $\mathcal{N}=1$ theory it is important to analyze the 
action of the gauge groups. 
The ansatz breaks the gauge group down to its abelian component. There are indeed 
eight residual $U(1)$ abelian gauge factors on the moduli space,
an $U(1)^2 $ in each $U(2)$.
They act on the $Q_{ij}$ as in the $\mathcal{N}=2$ case. We can check this explicitly as
follows.
\begin{eqnarray}
\mathcal{Q}_{12} &\rightarrow &
\left(\begin{array}{cc}e^{i\phi_1} & 0 \\0 &e^{-i\phi_8} \\ \end{array}\right)
\mathcal{Q}_{12}
\left(\begin{array}{cc}e^{i\phi_7} & 0 \\0 &e^{-i\phi_2} \\ \end{array}\right) 
\quad,\quad
\mathcal{Q}_{21} \rightarrow 
\left(\begin{array}{cc}e^{-i\phi_7} & 0 \\0 &e^{i\phi_2} \\ \end{array}\right)
\mathcal{Q}_{21}
\left(\begin{array}{cc}e^{-i\phi_1} & 0 \\0 &e^{i\phi_8} \\ \end{array}\right) 
\nonumber \\
\mathcal{Q}_{23} &\rightarrow &
\left(\begin{array}{cc}e^{-i\phi_7} & 0 \\0 &e^{i\phi_2} \\ \end{array}\right)
\mathcal{Q}_{23}
\left(\begin{array}{cc}e^{-i\phi_3} & 0 \\0 &e^{i\phi_6} \\ \end{array}\right) 
\quad,\quad
\mathcal{Q}_{32} \rightarrow 
\left(\begin{array}{cc}e^{i\phi_3} & 0 \\0 &e^{-i\phi_6} \\ \end{array}\right)
\mathcal{Q}_{32}
\left(\begin{array}{cc}e^{i\phi_7} & 0 \\0 &e^{-i\phi_2} \\ \end{array}\right) 
\nonumber \\
\mathcal{Q}_{34} &\rightarrow &
\left(\begin{array}{cc}e^{i\phi_3} & 0 \\0 &e^{-i\phi_6} \\ \end{array}\right)
\mathcal{Q}_{34}
\left(\begin{array}{cc}e^{i\phi_5} & 0 \\0 &e^{-i\phi_4} \\ \end{array}\right) 
\quad,\quad
\mathcal{Q}_{43} \rightarrow 
\left(\begin{array}{cc}e^{i\phi_5} & 0 \\0 &e^{-i\phi_4} \\ \end{array}\right)
\mathcal{Q}_{43}
\left(\begin{array}{cc}e^{i\phi_3} & 0 \\0 &e^{-i\phi_6} \\ \end{array}\right) 
\nonumber \\
\mathcal{Q}_{11} &\rightarrow &
\left(\begin{array}{cc}e^{i\phi_1} & 0 \\0 &e^{-i\phi_8} \\ \end{array}\right)
\mathcal{Q}_{11}
\left(\begin{array}{cc}e^{i\phi_1} & 0 \\0 &e^{-i\phi_8} \\ \end{array}\right) 
\quad,\quad
\mathcal{Q}_{44} \rightarrow 
\left(\begin{array}{cc}e^{-i\phi_4} & 0 \\0 &e^{i\phi_5} \\ \end{array}\right)
\mathcal{Q}_{44}
\left(\begin{array}{cc}e^{-i\phi_4} & 0 \\0 &e^{i\phi_5} \\ \end{array}\right) 
\nonumber\\
\end{eqnarray}
where $\phi_i$ are the phases of the $U(1)$s.
They are equivalent to
\begin{eqnarray}
Q_{12} &\rightarrow& e^{i(\phi_1-\phi_2)} Q_{12} ,\quad
Q_{21} \rightarrow e^{i(\phi_2-\phi_1)} Q_{21} ,\quad
Q_{23} \rightarrow e^{i(\phi_2-\phi_3)} Q_{23} ,\quad
Q_{32} \rightarrow  e^{i(\phi_3-\phi_2)} Q_{32} \nonumber \\
Q_{34} &\rightarrow& e^{i(\phi_3-\phi_4)} Q_{34} ,\quad
Q_{43} \rightarrow e^{i(\phi_4-\phi_3)} Q_{43} ,\quad
Q_{45} \rightarrow e^{i(\phi_4-\phi_5)} Q_{45} ,\quad
Q_{54} \rightarrow e^{i(\phi_5-\phi_4)} Q_{54}\nonumber \\
Q_{56} &\rightarrow& e^{i(\phi_5-\phi_6)} Q_{56} ,\quad
Q_{65} \rightarrow e^{i(\phi_6-\phi_5)} Q_{65},\quad
Q_{67} \rightarrow e^{i(\phi_6-\phi_7)} Q_{67} ,\quad
Q_{76} \rightarrow e^{i(\phi_7-\phi_6)} Q_{76}\nonumber \\
Q_{78} &\rightarrow& e^{i(\phi_7-\phi_8)} Q_{78} ,\quad
Q_{87} \rightarrow e^{i(\phi_8-\phi_7)} Q_{87},\quad
Q_{81} \rightarrow e^{i(\phi_8-\phi_1)} Q_{81} ,\quad
Q_{18} \rightarrow e^{i(\phi_1-\phi_8)} Q_{18} \nonumber \\
\end{eqnarray}
One can observe that one of them acts trivially, six combinations are
used to mod the moduli space and the last factor is broken to
$\mathbb{Z}_{2k}$ by the CS and consequently they reproduce exactly the
the $\mathcal{N}=2$ CY$_4$ geometry
(\ref{eq:geoCY41}) quotiented by the same $\mathbb{Z}_{2k}$ action.
Actually there is still a residual discrete
symmetry, $\Theta$, generated by $i \sigma_3$ in $U(2)$. It
corresponds to the antiholomorphic involution (\ref{actex444}) for which 
we need to mod out the geometry.  The moduli space is then 
the Spin$(7)$ quotient $Y/\Theta_k$, where $\Theta_k$ is the combination of
$\Theta$ with $\mathbb{Z}_{2k}$.  In this way the moduli space of the electric phase of the $\mathcal{N}=1$ 
theory is exactly the 
Spin$(7)$ geometry obtained by the anti-holomorphic involution on the
CY$_4$ of the parent $\mathcal{N}=2$ theory. 


\subsubsection*{Moduli space  of the magnetic $\mathcal{N}=1$ theory}

Here we study the moduli space for a single M2 brane in the dual
phase. Also in this case 
the gauge group for the $\mathcal{N}=1$ projected theory is the product of 
$U(2)$ factors, where, as before we disregarded the presence of additional fractional branes, 
that do not explore the moduli space. To solve the zero potential condition for the 
$\mathcal{N}=1$ theory we consider the ansatz for the scalar components of the $\mathcal{N}=1$ projected bifundamental 
fields
\bea &&
\mathcal{Q}_{12} = \phantom{-}
\frac{Q_{12}-Q_{87}^*}{2}\sigma_2-\frac{Q_{12}+Q_{87}^*}{2i}\sigma_1
\quad,\quad \mathcal{Q}_{21} = \phantom{-}
\frac{Q_{21}-Q_{78}^*}{2}\sigma_2+\frac{Q_{21}+Q_{78}^*}{2i}\sigma_1
\nonumber \\
&& \mathcal{Q}_{23} = \phantom{-}
\frac{Q_{23}-Q_{76}^*}{2}\sigma_2+\frac{Q_{23}+Q_{76}^*}{2i}\sigma_1
\quad,\quad \mathcal{Q}_{32} = -
\frac{Q_{32}+Q_{67}^*}{2}\sigma_2+\frac{Q_{32}-Q_{67}^*}{2i}\sigma_1
\nonumber \\
&& \mathcal{Q}_{34} = -
\frac{Q_{34}-Q_{65}^*}{2}\sigma_2+\frac{Q_{34}+Q_{65}^*}{2i}\sigma_1
\quad,\quad \mathcal{Q}_{43} = \phantom{-}
\frac{Q_{43}+Q_{56}^*}{2}\sigma_2+\frac{Q_{43}-Q_{56}^*}{2i}\sigma_1
\nonumber \\
\eea
and
\bea \mathcal{Q}_{11} = \frac{\sigma_1+i \sigma_2}{2}Q_{81}^* +
\frac{\sigma_1-i \sigma_2}{2}Q_{18} \quad,\quad \mathcal{Q}_{44} =
\frac{\sigma_1-i \sigma_2}{2}Q_{45}^* + \frac{\sigma_1+i
  \sigma_2}{2}Q_{54} 
\eea 
For the adjoints we have
%
\bea
\mathcal{X}_{11} = \frac{Q_{88}^*-Q_{11}}{2} I +
\frac{Q_{88}^*+Q_{11}}{2} \sigma_3 \quad,\quad \mathcal{X}_{33} =
\frac{Q_{33}^*+Q_{66}}{2} I - \frac{Q_{33}^*-Q_{66}}{2i} \sigma_3 \eea
By inserting this ansatz on the $\mathcal{N}=1$ superpotential
(\ref{WIIA}) we verified that the ansatz exactly reproduces the equations of motion (\ref{VB0}) of the
parent $\mathcal{N}=2$ theory.
Moreover we verified that ansatz exhausts the vacuum space.

To compute the moduli space we still need for the residual gauge symmetries.
There are eight residual $U(1)$ abelian gauge factors on the moduli
space, an $U(1)^2 $ in each $U(2)$.  They act as in the
$\mathcal{N}=2$ case. We can check this explicitly as follows:
\begin{eqnarray}
\mathcal{Q}_{12} &\rightarrow &
\left(\begin{array}{cc}e^{-i\phi_8} & 0 \\0 &e^{i\phi_1} \\ \end{array}\right)
\mathcal{Q}_{12}
\left(\begin{array}{cc}e^{-i\phi_2} & 0 \\0 &e^{i\phi_7} \\ \end{array}\right) 
\quad,\quad
\mathcal{Q}_{21} \rightarrow 
\left(\begin{array}{cc}e^{i\phi_2} & 0 \\0 &e^{-i\phi_7} \\ \end{array}\right)
\mathcal{Q}_{21}
\left(\begin{array}{cc}e^{i\phi_8} & 0 \\0 &e^{-i\phi_1} \\ \end{array}\right) 
\nonumber \\
\mathcal{Q}_{23} &\rightarrow &
\left(\begin{array}{cc}e^{i\phi_2} & 0 \\0 &e^{-i\phi_7} \\ \end{array}\right)
\mathcal{Q}_{23}
\left(\begin{array}{cc}e^{i\phi_6} & 0 \\0 &e^{-i\phi_3} \\ \end{array}\right) 
\quad,\quad
\mathcal{Q}_{32} \rightarrow 
\left(\begin{array}{cc}e^{-i\phi_6} & 0 \\0 &e^{i\phi_3} \\ \end{array}\right)
\mathcal{Q}_{32}
\left(\begin{array}{cc}e^{-i\phi_2} & 0 \\0 &e^{i\phi_7} \\ \end{array}\right) 
\nonumber \\
\mathcal{Q}_{34} &\rightarrow &
\left(\begin{array}{cc}e^{-i\phi_6} & 0 \\0 &e^{i\phi_3} \\ \end{array}\right)
\mathcal{Q}_{34}
\left(\begin{array}{cc}e^{-i\phi_4} & 0 \\0 &e^{i\phi_5} \\ \end{array}\right) 
\quad,\quad
\mathcal{Q}_{43} \rightarrow 
\left(\begin{array}{cc}e^{i\phi_4} & 0 \\0 &e^{-i\phi_5} \\ \end{array}\right)
\mathcal{Q}_{43}
\left(\begin{array}{cc}e^{i\phi_6} & 0 \\0 &e^{-i\phi_3} \\ \end{array}\right) 
\nonumber \\
\mathcal{Q}_{11} &\rightarrow &
\left(\begin{array}{cc}e^{i\phi_1} & 0 \\0 &e^{-i\phi_8} \\ \end{array}\right)
\mathcal{Q}_{11}
\left(\begin{array}{cc}e^{i\phi_1} & 0 \\0 &e^{-i\phi_8} \\ \end{array}\right) 
\quad,\quad
\mathcal{Q}_{44} \rightarrow 
\left(\begin{array}{cc}e^{-i\phi_4} & 0 \\0 &e^{i\phi_5} \\ \end{array}\right)
\mathcal{Q}_{44}
\left(\begin{array}{cc}e^{-i\phi_4} & 0 \\0 &e^{i\phi_5} \\ \end{array}\right) 
\nonumber \\
\mathcal{X}_{11} &\rightarrow &
\left(\begin{array}{cc}e^{i\phi_1} & 0 \\0 &e^{-i\phi_8} \\ \end{array}\right)
\mathcal{X}_{11}
\left(\begin{array}{cc}e^{-i\phi_1} & 0 \\0 &e^{i\phi_8} \\ \end{array}\right) 
\quad,\quad
\mathcal{X}_{33} \rightarrow 
\left(\begin{array}{cc}e^{-i\phi_3} & 0 \\0 &e^{i\phi_6} \\ \end{array}\right)
\mathcal{X}_{33}
\left(\begin{array}{cc}e^{i\phi_3} & 0 \\0 &e^{-i\phi_6} \\ \end{array}\right) 
\nonumber \\
\end{eqnarray}
and these are equivalent to the $\mathcal{N}=2$ action
\begin{eqnarray}
Q_{12} &\rightarrow& e^{i(\phi_1-\phi_2)} Q_{12} ,\quad
Q_{21} \rightarrow e^{i(\phi_2-\phi_1)} Q_{21} ,\quad
Q_{23} \rightarrow e^{i(\phi_2-\phi_3)} Q_{23} ,\quad
Q_{32} \rightarrow  e^{i(\phi_3-\phi_2)} Q_{32} \nonumber \\
Q_{34} &\rightarrow& e^{i(\phi_3-\phi_4)} Q_{34} ,\quad
Q_{43} \rightarrow e^{i(\phi_4-\phi_3)} Q_{43} ,\quad
Q_{45} \rightarrow e^{i(\phi_4-\phi_5)} Q_{45} ,\quad
Q_{54} \rightarrow e^{i(\phi_5-\phi_4)} Q_{54}\nonumber \\
Q_{56} &\rightarrow& e^{i(\phi_5-\phi_6)} Q_{56} ,\quad
Q_{65} \rightarrow e^{i(\phi_6-\phi_5)} Q_{65},\quad
Q_{67} \rightarrow e^{i(\phi_6-\phi_7)} Q_{67} ,\quad
Q_{76} \rightarrow e^{i(\phi_7-\phi_6)} Q_{76}\nonumber \\
Q_{78} &\rightarrow& e^{i(\phi_7-\phi_8)} Q_{78} ,\quad
Q_{87} \rightarrow e^{i(\phi_8-\phi_7)} Q_{87},\quad
Q_{81} \rightarrow e^{i(\phi_8-\phi_1)} Q_{81} ,\quad
Q_{18} \rightarrow e^{i(\phi_1-\phi_8)} Q_{18} \nonumber \\
Q_{11} &\rightarrow& \phantom{e^{i(\phi_1-\phi_8)}} Q_{11} ,\quad
Q_{33} \rightarrow \phantom{e^{i(\phi_1-\phi_8)}} Q_{33} ,\quad
Q_{66} \rightarrow \phantom{e^{i(\phi_1-\phi_8)}} Q_{66} ,\quad
Q_{88} \rightarrow \phantom{e^{i(\phi_1-\phi_8)}} Q_{88} \nonumber \\
\end{eqnarray}
One of them acts trivially, six combinations are used to mod the
moduli space and the last factor is broken to $\mathbb{Z}_{2k}$ by the
CS. We hence obtain exactly the CY$_4$ moduli space (\ref{CYp2}) quotiented
by the same $\mathbb{Z}_{2k}$ of the parent $\mathcal{N}=2$ theory.
Actually there is still a residual discrete symmetry, $\Theta$, generated
by $i \sigma_3$ in $U(2)$ that acts on the moduli space exactly as the 
anti-holomorphic involution (\ref{actex444}).
The moduli space is then the Spin$(7)$ quotient $Y/\Theta_k$, where
$\Theta_k$ is the combination of $\Theta$ with $\mathbb{Z}_{2k}$.  In
this way we computed the Spin$(7)$ geometry obtained by the
anti-holomorphic involution on the CY$_4$. It coincides with the
geometry of the other $\mathcal{N}=1$ theory introduced above.

The two Spin$(7)$ geometries $Y/\Theta_k$ computed by projecting the
toric dual parent theories coincide, and we conclude
that the two $\mathcal{N}=1$ models are Spin$(7)$ dual.
We conclude this section by arguing that in this case the Spin$(7)$ 
duality is actually a Seiberg-like duality.

\subsubsection*{Brane Construction and Seiberg-like Duality}

In this case we can support the duality between the two
$\mathcal{N}=1$ theories by using the brane construction.  One can
observe from figure \ref{fig:L444} that the orientifold projection in
this case folds the quiver by identifying pairs of $U(N)$ gauge
groups.  At the level of type IIB brane description the orientifolded theory is
locally $\mathcal{N}=2$. 
We can then exchange without problem the 
$(1,p_i)$ branes at the boundaries of the D3s associated to the 
second gauge group of the projected $\mathcal{N}=1$ theory. This operation generates the
Seiberg-dual phase exactly as in the parent theory, where the Seiberg duality is 
implemented at the same time on the two identified gauge groups. The $|p_i-p_{i+1}|$ fractional branes, created
during the exchange, modify the dual ranks $\widetilde N_2$, and we
have $\widetilde{N_2}=2(N+|k|)$. The CS levels transform as discussed
above and the superpotential transforms according to the usual rules
of Seiberg-like duality\footnote{We will come back to this issue in
  section \ref{sec:holodual}.}.  We conclude that in this case the Spin$(7)$
duality is a Seiberg-like duality.

\subsection{An infinite family}
\label{sec:gen}
\bef
\includegraphics[width=16cm]{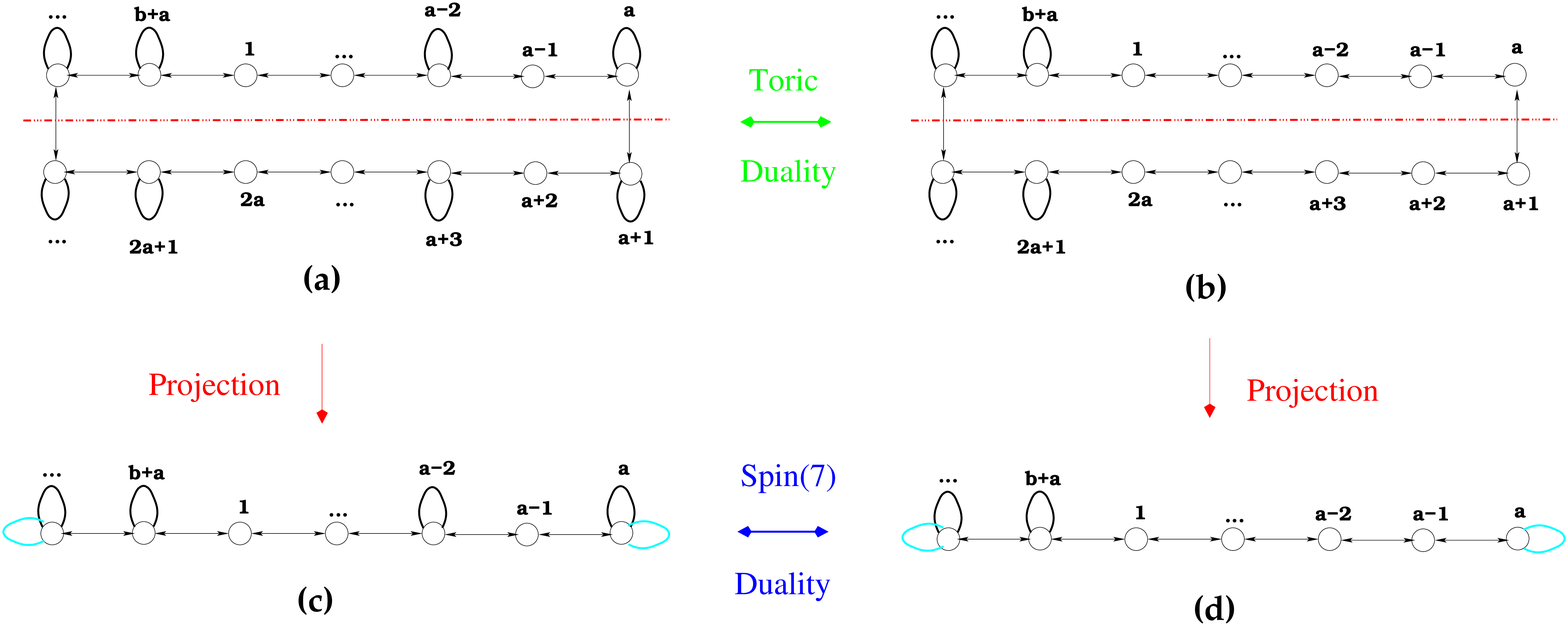}
\caption{
  The models in $(a)$ and $(b)$ represent two $L^{aba}$ toric dual phases
  with both $a$ and $b$ even.  The
  orientifold projection acts by folding the quiver, along the dashed
  red lines. By projecting these models to $\mathcal{N}=1$ Spin$(7)$
  cones we obtain the phases $(c)$ and $(d)$, that are related by
  Spin$(7)$ duality. These models are also Seiberg-like dual.}
\label{fig:gen}
\eef
In this section we propose a generalization of the Seiberg-like
duality discussed above, for an infinite family of $\mathcal{N}=1$
gauge theories.  The two dual phases are represented in figure
\ref{fig:gen} (c) and (d).  They can be obtained by projecting the
$\widetilde{L^{aba}}_{k_i}$ theories represented in figure \ref{fig:gen}
(a) and (b). We fix both $a$ and $b$ to be even.  We choose the CS
levels of the model in figure \ref{fig:gen} (a) as:
\begin{equation}
\left\{
\begin{array}{lll}
k_i = k &\quad\quad i=a-1& \quad\quad  i=a+2 \\
k_i = -k &\quad\quad   i=a&\quad\quad  i=a+1 \\
k_i = 0 &\quad\quad   \text{otherwise} &\\
\end{array}
\right.
\end{equation}
We can describe the geometry of these models in a unified way.  The
gauge invariant operators orthogonal to the CS vector are
\begin{eqnarray}
x_1 &=& Q_{12} Q_{21} = Q_{34} Q_{43} = \dots = Q_{2a-1,2a} Q_{2a,2a-1}=
Q_{2a+1,2a+1} = \dots =  Q_{b+a,b+a} 
\nonumber \\
x_2 &=& Q_{23} Q_{32} = Q_{45} Q_{54} = \dots = Q_{2a,2a+1} Q_{2a+1,2a+1}=
Q_{2a+1,2a+2}Q_{2a+2,2a+1} = \dots  \nonumber \\
\dots &=&  Q_{b+a-1,b+a} Q_{b+a,b+a-1}=
Q_{b+a,1} Q_{1,b+a}
\nonumber\\
y_1 &=& Q_{12} Q_{23} \dots Q_{b+a,1}, \quad
y_2 = Q_{1,b+a} Q_{b+a,b+a-1}\dots Q_{21} \nonumber \\
t_1 &=& Q_{a-1,a}Q_{a+2,a+1}, \quad
t_1 = Q_{a+1,a+1}Q_{a,a-1}
\end{eqnarray}
The CY$_4$ $Y$ geometry that has to be mod by the $\mathbb{Z}_k$ is
\begin{equation}
y_1 y_2 = x_1^a x_2^b \quad,\quad t_1 t_2 = x_2^2
\end{equation}
We choose the anti-holomorphic involution by generalizing the choice
of section \ref{es:2}.
On the $Y$ coordinates it is
\begin{equation}
x_1 \rightarrow x_1^*\quad 
x_2 \rightarrow -x_2^*\quad 
y_1\rightarrow y_2^*\quad 
y_2 \rightarrow y_1^*\quad 
t_1 \rightarrow
t_1^*\quad 
t_2 \rightarrow t_2^* 
\end{equation}
This action has fixed points and it can be translated on the vector
and matter multiplets as usual. It sends $\sigma \rightarrow \sigma$
and $\psi \rightarrow -\psi$ and it hence acts as an
orientifold projection on the quiver field theory.
It can indeed be realized quotienting by 
an antiholomorphic orientifold symmetry of the $\mathcal{N}=2$ lagrangian that identifies
$i$-th group with the $i+(a+b)/2$-th. The rank of each
gauge group and the CS level are doubled after the identification.
Finally we obtain the $\mathcal{N}=1$ theory represented in figure
\ref{fig:gen} (c).

We choose to dualize this family by acting on nodes $a-1$ and
$a+2$. Other choices are possible. The quiver is depicted in figure
\ref{fig:gen} (b).  The new CS levels are
\begin{equation}\label{newcs}
\left\{
\begin{array}{lll}
k_i = -k &\quad\quad i=a-1& \quad\quad  i=a+2 \\
k_i = -k &\quad\quad   i=a-2&\quad\quad  i=a+3 \\
k_i = 0 &\quad\quad   \text{otherwise} &
\end{array}
\right.
\end{equation}
This theory has the same CY$_4$ geometry $Y$ as before.
One can verify that, by implementing the same anti-holomorphic
involution on the coordinates as before, we obtain the quiver in figure
\ref{fig:gen} (d). 

We computed the Spin$(7)$ geometries $Y/\Theta_k$ of both the $\mathcal{N}=1$
models and we showed that they coincide.  This confirms that they are
Spin$(7)$ dual.  As in the subsection \ref{es:2} the Spin$(7)$
duality is also in this case a Seiberg-like duality. Indeed one can
embed the $\mathcal{N}=1$ theories in a IIB brane setup and observe
that locally the Seiberg-like duality can be performed ignoring the
effect of the orientifold.
%
%
%
%
%
%
%
%
%
%
%
%
%
%
%
%
%
%
\section{Comments on $\mathcal{N}=1$ Seiberg-like duality }
\label{sec:holodual}

As discussed in the previous section in some cases we can claim that
the proposed Spin$(7)$ duality coincides with three dimensional
Seiberg-like duality for $\mathcal{N}=1$ theories.  We used a brane
description to support this idea.

In this section we translate this brane description in a field
theoretical language. We propose a procedure to obtain the
Seiberg-like dual in the $\mathcal{N}=1$ case. We provide the
transformation rules on the superpotential, on the field content and
on the gauge groups by extracting them from the example in subsection
\ref{es:2}.

We can summarize the procedure as follows.  First the gauge invariant
operators of the electric theory appear as mesons in the magnetic
theory. Then the ranks of the gauge groups and the CS levels transform
as in (\ref{N2r}).  Moreover we distinguish three terms in the dual
$\mathcal{N}=1$ superpotential. The first term is a holomorphic
contribution that we call $W_{holo}$. It is cubic and involves the
coupling of the dual quarks with the mesons.  The second part is non
holomorphic, and it is obtained from the $\mathcal{N}=1$
superpotential after a proper substitution of the electric quarks with
the mesons. In the case of the Spin$(7)$ duality this term corresponds
to the $\mathcal{N}=2$ superpotential of the parent theory, projected
to $\mathcal{N}=1$.  The last term is obtained by coupling the dual
$R$ fields with the D terms.  The masses of the $R$ fields are
proportional to the dual CS levels.

In the first part of the section we show that these rules reproduce
the dual theory studied in subsection \ref{es:2}.  Then we apply this
procedure to one $\mathcal{N}=1$ theory with $U(N_c)_k$ gauge group,
$N_f$ flavors and a quartic, non holomorphic, superpotential. This
theory is not associated to a CY$_4$ and we cannot use the Spin$(7)$
duality. In any case we propose a possible Seiberg-like dual
description. We obtain a dual $U(N_f+|k|-N_c)_{-k}$ gauge theory with
$N_f$ flavors and the same quartic, non holomorphic, superpotential.
%
%
%
%
%
%
%
%
%
%
%
%
%
%
%
%
%
%
\subsection{Revisiting  the $\widetilde L^{444}_{k_i}$ theory}
\label{subsec:revisiting}
Here we  reconsider the models studied in section \ref{es:2}.
We start from the electric theory, with superpotential (\ref{WIA}). 

Here we follow the procedure sketched above to obtain the dual phase.
First we identify the group to dualize.  This group and its
neighbours are modified as (\ref{N2r}).  Then we can build the
dual superpotential.  We start from the holomorphic term:
\begin{eqnarray}
\label{holoD}
W_{holo}&=&
\mathcal{X}_{11} \mathcal{Q}_{12} \mathcal{Q}_{21} 
+ \mathcal{X}_{33} \mathcal{Q}_{32}\mathcal{Q}_{23}
+\mathcal{X}_{13} \mathcal{Q}_{32} \mathcal{Q}_{21} 
+ \mathcal{X}_{31} \mathcal{Q}_{12}\mathcal{Q}_{23}
+h.c.
\end{eqnarray}
The next step consists of contracting the fields charged under the
dualized gauge groups into mesons $\mathcal{X}_{ij}$. They are
\begin{equation}
\left(
\begin{array}{cc}
\mathcal{X}_{11}&\mathcal{X}_{13}\\
\mathcal{X}_{31}&\mathcal{X}_{33}
\end{array}
\right)
=
\left(
\begin{array}{cc}
Q_{12} Q_{21} & Q_{12} Q_{23}  \\
Q_{32} Q_{21} & Q_{32} Q_{23} 
\end{array}
\right)
\end{equation}
We substitute the mesons in the first two lines of (\ref{WIA}) and
integrate them out the massive fields.  This procedure reproduces the
first two lines of (\ref{WIIA}).  The other terms of (\ref{WIIA}) are
obtained by reintroducing the $D$-terms and the $R$ fields.  The mass
terms for $R_i$ are obtained by transforming the CS levels with the
usual rule (\ref{N2r}).  In this way we reproduced the dual theory
discussed in section \ref{es:2}.
%
%
%
%
%
%
%
%
%
%
\subsection{$U(N)_k$ SQCD with quartic superpotential}
\label{subsec:proposal}

Even if we derived the rules of the $\mathcal{N}=1$ Seiberg-like
duality from a specific set of theories, describing M2 branes probing
Spin$(7)$ singularities, we can try to push further in the field
theoretical direction. Here we apply these rules to a SQCD like
model, that does not have a known AdS$_4$ dual. 
We propose a dual version of $U(2N_c)_{2k}$ SQCD with
$2N_f$ flavors and a non-holomorphic quartic superpotential
\begin{equation}
\label{WeleNcNfk}
W = Q \tilde Q Q^* \tilde Q^* + \frac{k}{2\pi} R^2
+ R \left(Q Q^{\dagger} - \widetilde Q^\dagger \widetilde Q
\right) 
\end{equation}  
We study the dual of the non-holomorphic superpotential
(\ref{WeleNcNfk}) and the quarks are $\mathcal{N}=1$ complex scalar
superfields.  The dual theory is obtained by applying the rules
explained above in the case of the quiver gauge theories.  The dual
gauge group is expected to be $U(2N_f-2N_c+2|k|)_{-2k}$.  There are
$2N_f$ dual flavors and the meson $M = Q \tilde Q$.  The holomorphic
part of the dual superpotential is
\begin{equation}
  W_{holo} = M q \tilde q +  M^* q^* \tilde q^* 
\end{equation}
By considering the deformation $ Q \tilde Q Q^* \tilde
Q^* = M M^*$ we can integrate out the meson $M$.  By turning on the
contributions of the $D$ terms and of the $R$ field, with $k_{CS}=-2k$ we
have
\begin{equation}
W = q \tilde q  q^* \tilde q^*  -\frac{k}{2\pi} \widetilde{R}^2
+ 
\widetilde R \left( \widetilde q   \widetilde q^{\dagger} 
- q^\dagger q
\right) 
\end{equation}
As in the quartic $\mathcal{N}=2$ SQCD in three dimensions or the
quartic $\mathcal{N}=1$ SQCD in four dimensions we observe that our
procedure predicts the self duality for $\mathcal{N}=1$ three
dimensional CS-SQCD with a quartic interaction.  Anyway, in general,
this theory is not superconformal and moreover it is not protected
against quantum corrections. It would be interesting to provide some
checks of this duality, by engineering a brane realization and by
computing the Witten index, by first lifting the moduli space
in both phases consistently. We
leave this analysis to future investigations.
%
%
%
%
%
%
%
%
%
%
\section{Discussion and further developments}
\label{sec:conclusions}

In this paper we proposed a generalization of $\mathcal{N}=2$ toric
duality for M2 branes probing toric CY$_4$ singularities to
$\mathcal{N}=1$ models of M2 branes probing Spin$(7)$
singularities. We called this generalization Spin$(7)$ duality. This
proposal has been supported by the AdS/CFT correspondence. Indeed we
 matched the moduli space of $\mathcal{N}=1$ Spin$(7)$ dual models
by orientifolding $\mathcal{N}=2$ toric dual pairs. In some cases, with the help
of the brane picture, we argued that the Spin$(7)$ duality is also a
Seiberg-like duality.  Finally we proposed a generalization of this
$\mathcal{N}=1$ Seiberg-like duality for models without a known AdS
dual description.

The main problem in the study of a supersymmetric, but non
holomorphic, duality is its validity at quantum level. In the near
horizon limit the AdS/CFT correspondence provides some arguments 
to protect the validity of the duality beyond the classical level.
The 
strongly coupled phases of the dual pair of theories are conjectured
to describe the QFT of M2 branes probing the same Spin$(7)$ cone. By
considering the near horizon geometry the models are superconformal
invariant and represent two dual descriptions of the same singularity that 
should hence be valid in the strong coupling region. Planar equivalence 
moreover 
supports the duality between the
pairs for large N.

However other checks are necessary.  For example one should compute the Witten
index \cite{Witten:1999ds,Smilga:2009ds,Smilga:2013usa} to match the
number of supersymmetric vacua.  Moreover it would be interesting to
study other partition functions, by localizing the $\mathcal{N}=1$
models on more complicate manifolds, like the three sphere $S^3$ .
In the $\mathcal{N}=2$ case \cite{Jafferis:2010un,Hama:2010a} toric
duality on the three sphere has been checked for the
$\widetilde{L^{aba}}_{k_i}$ theories in \cite{Martelli:2011qj,
  Jafferis:2011zi, Amariti:2011uw, Amariti:2012tj}. A generalization
of this analysis to Seiberg-like duality for these theories appeared
in \cite{Agarwal:2012wd}.  In the $\mathcal{N}=1$ case the
calculations may be very involved, because of the absence of
holomorphy and of a continuous $R$-symmetry, but they can potentially
provide strong checks of the dualities.

Another interesting aspect regards other possible models. Here we
discussed only vector like models, but there are also $\mathcal{N}=2$
chiral models with an AdS$_4$ dual \cite{Benini:2009qs}.  They correspond
to quiver gauge theories with vector-like bifundamentals and chiral
flavors. It should be interesting to extend the orientifold projection
to these models and study the Spin$(7)$ duality for those cases.

\section*{Acknowledgements}

We would like to thank Diego Redigolo for his valuable participation
in the early stage of this project and for many enlightening
discussions.  It is a great pleasure to thank Alberto Zaffaroni for
comments on the draft.  We are also grateful to Costas Bachas, Cyril
Closset, Amihay Hanany, Ken Intriligator, Dan Israel and Jan Troost for
discussions and comments.  D.F would like to acknowledge the kind
hospitality of the LPTHE, where part of this research has been
implemented.  A.A. is grateful to the Institut de Physique Th\'eorique
Philippe Meyer at the \'Ecole Normale Sup\'erieure for fundings.
D.F. is a ``Charg\'e de recherches'' of the Fonds de la Recherche
Scientifique--F.R.S.-FNRS (Belgium), and his research is supported by
the F.R.S.-FNRS and partially by IISN - Belgium (conventions 4.4511.06
and 4.4514.08), by the ``Communaut\'e Fran\c{c}aise de Belgique''
through the ARC program and by the ERC through the ``SyDuGraM''
Advanced Grant.

\appendix
%
%
%
%
%
%
%
%
%
%
%
%
%
%
%
%
%
\section{$\mathcal{N}=1$ formalism}
\label{APPA}
In this appendix we quickly report some known results about the
$\mathcal{N}=1$ superspace obtained by setting to zero some of the
Grassmann variables of the $\mathcal{N}=2$ case in three dimensions.

First we start by reviewing the $\mathcal{N}=2$ case.  There are two
possible multiplets involved in a $U(N)$ quiver, the vector multiplet and
the bifundamental chiral multiplet.  In a quiver with $G$ nodes the
a-th vector multiplet $V_a$ contains a three dimensional gauge field
$A_\mu$, a two component Dirac spinor and two real scalars, $\sigma_a$
and $D_a$.  A chiral bifundamental $X_{ab}$ connecting the $a$-th and
the $b$-th node (if $a=b$ we have a chiral field in the adjoint
representation) than consists of two complex scalars and a two
dimensional Dirac spinor.  The $\mathcal{N}=1$ superspace is obtained
by decomposing the $\mathcal{N}=2$ case in two copies of
$\mathcal{N}=1$ and by projecting out one of them
\cite{Gates:1983nr,Avdeev:1991za,Avdeev:1992jt}. The decomposition
is obtained by splitting the $\theta$ variables and the
super-derivatives as
\begin{equation}
\theta_{\alpha} = \theta_{1\alpha} + i  \theta_{2\alpha}
\quad,\quad
D_{\alpha} = \frac{1}{2}\left(D_{1\alpha} + i D_{2\alpha}\right)
\quad,\quad
\overline D_{\alpha} = \frac{1}{2}\left(D_{1\alpha} - i D_{2\alpha}\right) 
\end{equation}
and the projection to $\mathcal{N}=1$ is performed by setting
$\theta_2=0$ in the lagrangian.  In terms of superfields the
$\mathcal{N}=2$ vector multiplet decomposes into a $\mathcal{N}=1$
spinor superfield $\Gamma_{\alpha}^a$ and an $\mathcal{N}=1$ auxiliary
real scalar superfield $R_a$.  The chiral multiplet $X_{ab}$
decomposes into two $\mathcal{N}=1$ real scalar superfields,
$Re(X_{ab})$ and $Im(X_{ab})$ that can be combined into a single
complex scalar superfield $Y_{ab}$.

By acting on the $\mathcal{N}=2$ lagrangian the $\mathcal{N}=1$
superpotential has three different contributions. They are
\begin{eqnarray}
\frac{k_a}{8 \pi} S_{CS_a}^{\mathcal{N}=2} &\rightarrow& -\frac{k_a}{4 \pi} \int d^2 \theta_1 R_{a}^2 
\nonumber \\
-\int d^4 \theta X_{ab}^{\dagger}e^{-V_a}X_{ab}e^{V_a}&\rightarrow& \int d^2\theta_1 \left(Y_{ab}Y_{ab}^{\dagger}R_a - Y_{ab}^\dagger Y_{ab}R_b\right)
\nonumber \\
\int d^2 \theta W(X_{ab} + c.c.  &\rightarrow&  \int d^2 \theta_1 \left(W(Y_{ab})+W(Y_{ab}^*)\right)
\nonumber
\end{eqnarray}
%
%
%
%
%
%
%
%
%
%
%
%
%
%
%
%
%
\section{$\mathcal{N}=1$ superconformal algebra}
\label{APPB}
In this appendix we provide the generic structure of the
superconformal algebra in three dimensional $\mathcal{N}=1$
theories  \cite{Nahm:1977tg,Park:1999cw}.

We define the two dimensional Gamma matrices $\gamma^{\mu}$, $\mu=0,1,2$,
satisfying the relations
\begin{eqnarray}
\gamma^\mu \gamma^\nu &=& 
\eta ^{\mu \nu} + i \epsilon^{\mu \nu \rho} \gamma_\rho
\end{eqnarray}
with $\eta^{\mu \nu} = (1,-1,-1)$.
The  three dimensional $\mathcal{N}=1$ superconformal algebra is
\begin{equation}
\begin{array}{crcl} 
  & [P_\mu,P_\nu] &=& [P_{\mu},Q]= [k_\mu,k_\nu]=[D,M_{\mu \nu}]=[D,D] =[K_\mu,S] =0   \\ 
  & [M_{\mu \nu},P_\lambda ] &=& i(\eta_{\mu \nu}P_\nu - \eta_{\nu \lambda} P_\mu) ,\quad 
   [M_{\mu \nu},M_{\lambda \rho} ]=i
  \left (\eta_{\mu \lambda}M_{\nu \rho} -
    \eta_{\mu \rho} M_{\nu \lambda}
    -\eta_{\nu \lambda} M_{\mu \rho}
    +\eta_{\nu \rho} M_{\mu \lambda} \right),  \\ 
  & \{ Q,Q \} &=& 2 \gamma^\mu P_\mu   ,\quad
   [M_{\mu \nu},Q] = \frac{i}{2}\gamma_{[\mu}\gamma_{\nu]}Q   ,\quad
    [M_{\mu \nu}, K_\lambda]=i(\eta{\mu \lambda} K_{\nu} - \eta_{\nu \lambda} K_\mu),   \\
  & \{S,S\} &=& 2 \gamma^\mu K_\mu  ,\quad
   [M_{\mu \nu},S] = \frac{i}{2} \gamma_{[\mu}\gamma_{\nu]} S  ,\quad
   [P_\mu,K_\nu] = 2 i (M_{\mu \nu}+\eta_{\mu \nu} D),   \\
  & [P_\mu,S] &=& -\gamma_\mu Q   ,\quad
   [K_\mu,Q] = -\gamma_\mu S   ,\quad
   \{ Q,S \} = -i\left(2 D + \gamma^{[\mu}\gamma^{\nu]} M_{\mu \nu}\right),   \\
  & [D,P_\mu] &=& - i P_\mu,   \quad
    [D,S] =\frac{i}{2} S,   \quad
    [D,K_\mu] = i K_\mu ,  \quad
    [D,Q] = -\frac{i}{2} Q    
\end{array}
\end{equation}

\bibliographystyle{JHEP}
\bibliography{BibFile}

\end{document}